\documentclass[a4paper,11pt]{article}
\pdfoutput=1

\usepackage{jheppub}
\usepackage{graphics, setspace}

\newcommand{\mathsym}[1]{{}}
\newcommand{\unicode}[1]{{}}

\usepackage{amsmath}
\usepackage{amssymb}
\usepackage{bbold}
\usepackage{xspace}
\usepackage{verbatim}
\usepackage{bm}

\usepackage{subfig}
\usepackage[countmax]{subfloat}
\usepackage{slashed}

\usepackage{comment}

\usepackage{bbm}

\newcommand{\nbar}{{\bar n}}

\newcommand{\ecf}[2]{e_{#1}^{(#2)}}

\newcommand{\ecflp}[2]{\tilde e_{#1}^{(#2)}}
\newcommand{\ecfop}[2]{\mathbf{E_{#1}}^{(#2)}}

\newcommand{\sja}{n_{sj}}
\newcommand{\sjabar}{\bar{n}_{sj}}
\newcommand{\outj}{\tau}
\newcommand{\outjlp}{\tilde \tau}

\def\be{\begin{equation}}
\def\ee{\end{equation}}

\def\nbar{\bar n}

\newcommand{\ecfLa}{e_{2}^{(\alpha)}}

\newcommand{\ecfres}{e_{3}^{(\alpha)}}

\newcommand{\sjtheta}{\theta_{sj}}

\DeclareRobustCommand{\Sec}[1]{Sec.~\ref{#1}}

\DeclareRobustCommand{\App}[1]{App.~\ref{#1}}

\DeclareRobustCommand{\Fig}[1]{Fig.~\ref{#1}}

\DeclareRobustCommand{\Eq}[1]{Eq.~(\ref{#1})}
\DeclareRobustCommand{\Eqs}[2]{Eqs.~(\ref{#1}) and (\ref{#2})}
\DeclareRobustCommand{\Ref}[1]{Ref.~\cite{#1}}
\DeclareRobustCommand{\Refs}[1]{Refs.~\cite{#1}}

\newcommand{\vincia}[1]{\textsc{Vincia\xspace #1}}

\newcommand{\ariadne}{\textsc{Ariadne}}

\usepackage{color}

\preprint{MIT-CTP/4713}

\title{The Edge of Jets and Subleading Non-Global Logs}

\author{Duff Neill}

\affiliation{Center for Theoretical Physics, Massachusetts Institute of Technology, Cambridge, MA 02139, USA}

\emailAdd{dneill@mit.edu}

\abstract{
A persistent and fascinating problem at the high energy colliders are jets. Often trying to observe physics underlying the hard interactions at colliders requires experimental cuts in phase space, defining several jet or beam regions. QCD being a gauge theory that readily decays into infra-red modes, correlations between jet regions is almost inevitable, spoiling the predictivity of fixed order QCD calculations. One is faced with the task of calculating the evolution of a reduced density matrix, where successively less energetic (jet) regions are integrated out, to gain control of the calculation. I relate the decay rates governing the flow into the IR to an effective field theory expansion in soft jets, allowing a systematic and resummed calculation of these rates, while further relating them to physically observable features of the QCD cascade. To demonstrate the utility of the soft jet expansion, I present a factorization theorem for a soft subjet collinearly splitting in and out of a parent fat jet.  Using the resummation properties of this factorization theorem, I elucidate the structure of the subleading non-global logs (encoding the jet correlations) in the hemisphere jet mass distribution, as well as give a collinear improvement of the leading order resummation equation, the BMS equation. I compare to other approaches to subleading resummation of NGLs, and find the collinear improvement of the leading order equation removes the need for kinematic-dependent corrections in the IR averaging procedure of the reduced density matrix, so that no further large logs can be generated in the IR. Finally I end with speculation about connections with collinear improvements of the NLO B-JIMWLK hierarchy for small-$x$ resummation.
}

\begin{document} 
\maketitle

\section{Introduction}
Non-global logarithms (NGLs) pose an fascinating problem for resummation \cite{Dasgupta:2001sh}, and can have a large phenomenological impact \cite{Banfi:2010pa, Dasgupta:2012hg}. These are logarithms associated with soft radiation which splits and lands into different angular phase space regions, with an energy hierarchy between the soft daughters. Physically one pictures as an initial condition multiple hard jets at wide fixed angles in the event, whose masses are all much smaller than the initial hard collision. However, the masses themselves exhibit a hierarchy of scales. The resummation is difficult, since the soft radiation is sensitive to the pattern of previous soft emissions. This is in contrast to global logarithms, the logarithms of the jet mass to the hard scale, which are associated with virtual corrections of QCD amplitudes. For the global corrections, no parton becomes a physical state between the hard interaction scale, and the jet mass scale, so that while the cancellation of IR divergences between real and virtual corrections still occurs, a large log is left over. The exponentiation of these global logarithms is now a standard technique, whether using factorization theorems \cite{Kidonakis:1997gm,Kidonakis:1998bk,Berger:2003iw, Fleming:2007qr,Fleming:2007xt,Schwartz:2007ib,Bauer:2008dt,Hornig:2009vb,Ellis:2010rwa}, or QCD coherence arguments \cite{Catani:1990rr,Catani:1991bd,Catani:1992ua,Dokshitzer:1998kz,Banfi:2004yd}, to the point that it can be essentially automated \cite{guido_talk,Banfi:2014sua,Gerwick:2014gya,Farhi:2015jca} in both approaches. Moreover, the exponentiation is truly an explicit exponentiation, at least in an appropriate conjugate space for additive observables. However, the fixed order pattern of NGLs as calculated in \cite{Schwartz:2014wha} and \cite{Khelifa-Kerfa:2015mma} do not exhibit any such straightforward pattern.

Dealing with the resummation of NGLs, four approaches have appeared in the literature: avoiding them entirely with clever observables \cite{Dasgupta:2013via, Dasgupta:2013ihk,Larkoski:2014wba}, a large-$N_c$ Monte Carlo technique \cite{Dasgupta:2001sh}, non-linear evolution equations\footnote{For NGLs, this evolution equation is called the BMS equation after its inventors, Banfi, Marchesini, and Smye. However, it fits into a universality class of quantum master equations, see \Sec{sec:Lindblad} for a discussion on the universality of IR evolution equations.} \cite{Banfi:2002hw,Weigert:2003mm,Marchesini:2003nh,Marchesini:2004ne, Hatta:2008st,Avsar:2009yb,Hatta:2013iba,Hagiwara:2015bia}\footnote{For progress towards an EFT interpretation of these evolution equations, see \Ref{Becher:2015hka}, which incorporates an object akin to the trace of the post-evolution reduced density matrix of \Ref{Caron-Huot:2015bja} into their jet factorization theorem.}, and an expansion in soft sub-jets \cite{Forshaw:2006fk,Forshaw:2008cq,Forshaw:2009fz,DuranDelgado:2011tp,Larkoski:2015zka}. This last approach can be grounded in the structure of factorization theorems, as derived in the framework of soft collinear effective field theory (SCET) \cite{Bauer:2000yr, Bauer:2001ct, Bauer:2001yt}, promising a straightforward way to calculate to higher orders in $\alpha_s$ for the anomalous dimensions in the resummation. Perhaps even more importantly, \Ref{Larkoski:2015zka} organized the phase space of emissions into parametrically separated regions where distinct (now global) resummations apply. This is accomplished by using a complete set of infra-red and collinear safe observables, for example see \Refs{Stewart:2010tn, Larkoski:2013eya}, to define multi-differential cross-sections with sufficient number of observables to distinguish the soft and collinear limits of multiple pronged jet structures \cite{Bauer:2011uc,Larkoski:2014tva,Larkoski:2014gra,Larkoski:2014zma,Procura:2014cba}. Resumming these cross-sections, and expressing the more inclusive cross-sections as a marginalization over the more exclusive ones, the perturbation series is re-organized as summing over jets (resummed partons), rather than fixed order partons. Interesting emergent behavior, like the buffer region \cite{Dasgupta:2002bw}, manifests itself for these dressed emissions. Since the observables are complete up-to a resolution scale (which terminates the number of jets included in the calculation), one can also check whether all phase space regions are accounted for by examining all possible relative scalings of the observables. This has important implications for the subleading NGLs, since the fixed order NGLs at $\alpha_s^2$ in QCD are sensitive to a distinct phase space region, where a soft subjet is collinearly splitting along the jet boundary. This was already noted in \Ref{Larkoski:2015zka}. Here I shall worked out the soft jet factorization in this region, and its implications for resumming NGLs. 

The organization of this paper is as follows: I set out the phase space region for a soft subjet collinearly splitting at the jet boundary. Or more concisely, I consider edge of jet sub-jets. Having worked out the phase space region, I then give the appropriate factorization theorem for it. I then calculate to one-loop the objects found in the factorization theorem, finding both double logarithms of collinear origin, as well as a DGLAP style splitting process \cite{Gribov:1972ri,Dokshitzer:1977sg,Altarelli:1977zs}. Having these tools in hand, I show how the single dressed gluon with these collinear effects can be included to resum NGLs of more inclusive observables, following the procedure of \cite{Larkoski:2015zka}. I then compare to the NLO BMS equation derived in \cite{Caron-Huot:2015bja}. I find that this evolution equation naively misses the fixed order $\alpha_s^2$ subleading NGLs, which would be corrected with the appropriate NLO calculation for the trace of the density matrix. I find this to be connected with the lack of collinear evolution in the leading order BMS equation. Going back to the original BMS equation, and using the soft jet factorization theorem to calculate soft jet production at large-$N_c$, I show how one can collinearly improve the equation to include these collinear double logarithms. Then I discuss whether one can find any other IR structure beyond out of jet fragmentation. Then I argue that the soft jet expansion gives a systematic view of calculating the resummed decay rates used in these universal IR evolution equations, highlighting its physically observable features. Finally, having started at the beginning, I then conclude at the end with speculations on small-$x$ physics. 

\section{Phase Space Region}
For the purpose of understanding the collinear effects in subleading NGLs, I consider $e^+e^-\rightarrow hadrons$ with the hard scale $Q^2$, and wish to isolate the region of phase space associated with a soft subjet with two hard jets. In particular, I suppose two hard jet regions, the ``out-of-fat-jet'' region where an inclusive jet shape measurement has been made, and a fat jet region containing (most) of the soft subjet. The two hard jet regions are back to back, with associated light cone vectors:
\begin{align} 
&\text{Fat Jet Axis: } n=(1,\hat{n}),\\
&\text{Recoiling Jet Axis: }\nbar=(1,-\hat{n}),\\
&n\cdot\nbar=2,\qquad n^2=\nbar^2=0\,.
\end{align}
The soft subjet enjoys its own light-cone direction $\sja=(1,\hat{n}_{sj})$ with conjugate $\sjabar=(1,-\hat{n}_{sj})$. 

The phase space region for collinear splittings along the jet boundary exists strictly only for cone algorithms. A recombination algorithm will necessarily deform the jet boundary if a soft jet is present, though this effect can by mitigated with the anti-k$_t$ algorithm \cite{Cacciari:2008gp}, where the factorization of \cite{Larkoski:2015zka} also applies. Here I suppose that the jet axis is defined using the broadening \cite{Larkoski:2014uqa} or thrust axis \cite{Stewart:2010tn} of the event, and a fixed cone of radius $R$ is drawn about the jet (this is similar to the set up in \Ref{Kelley:2011aa}). For multi-jet events, an ideal cone algorithm for this factorization theorem is the XCone algorithm of \Refs{Stewart:2015waa,Thaler:2015xaa}.  

Within the fat jet, I measure the energy correlation functions $\ecf{2}{\alpha}, \ecf{2}{\beta},$ and $\ecf{3}{\beta}$ \cite{Larkoski:2013eya} to isolate the appropriate region of phase space\footnote{For a complete and detailed discussion of the $1\rightarrow 2$ phase space for QCD radiation, and the impact the different regions have on the form of factorization, please see \cite{Larkoski:2015kga}.}, whose definitions are summarized in \App{app:ecfs}. In this cross-section, some out-of-fat-jet measurement has been made, $\outj$. I shall for definiteness focus on the out-of-jet thrust measurement, with its action on a state $|X\rangle$ given by:
\begin{align}
\Theta_{\overline{FJ}}\mathbf{\outj}|X\rangle &=\sum_{i\in X}\nbar\cdot p_i\theta_{\overline{FJ}}(p_i)|X\rangle
\end{align}
The jet algorithm constraints on a single particle state are:
\begin{align}\label{eq:fat_jet_constraint}
\Theta_{FJ}|p\rangle&=\theta_{FJ}(p)|p\rangle=\theta\Big(\text{tan}^2\frac{R}{2}-\frac{n\cdot p}{\nbar\cdot p}\Big)|p\rangle\nonumber\\
\Theta_{\overline{FJ}}|p\rangle&=\theta_{\overline{FJ}}(p)|p\rangle=\theta\Big(-\text{tan}^2\frac{R}{2}+\frac{n\cdot p}{\nbar\cdot p}\Big)|p\rangle
\end{align}

\begin{figure}
\begin{center}
\includegraphics[width=6cm]{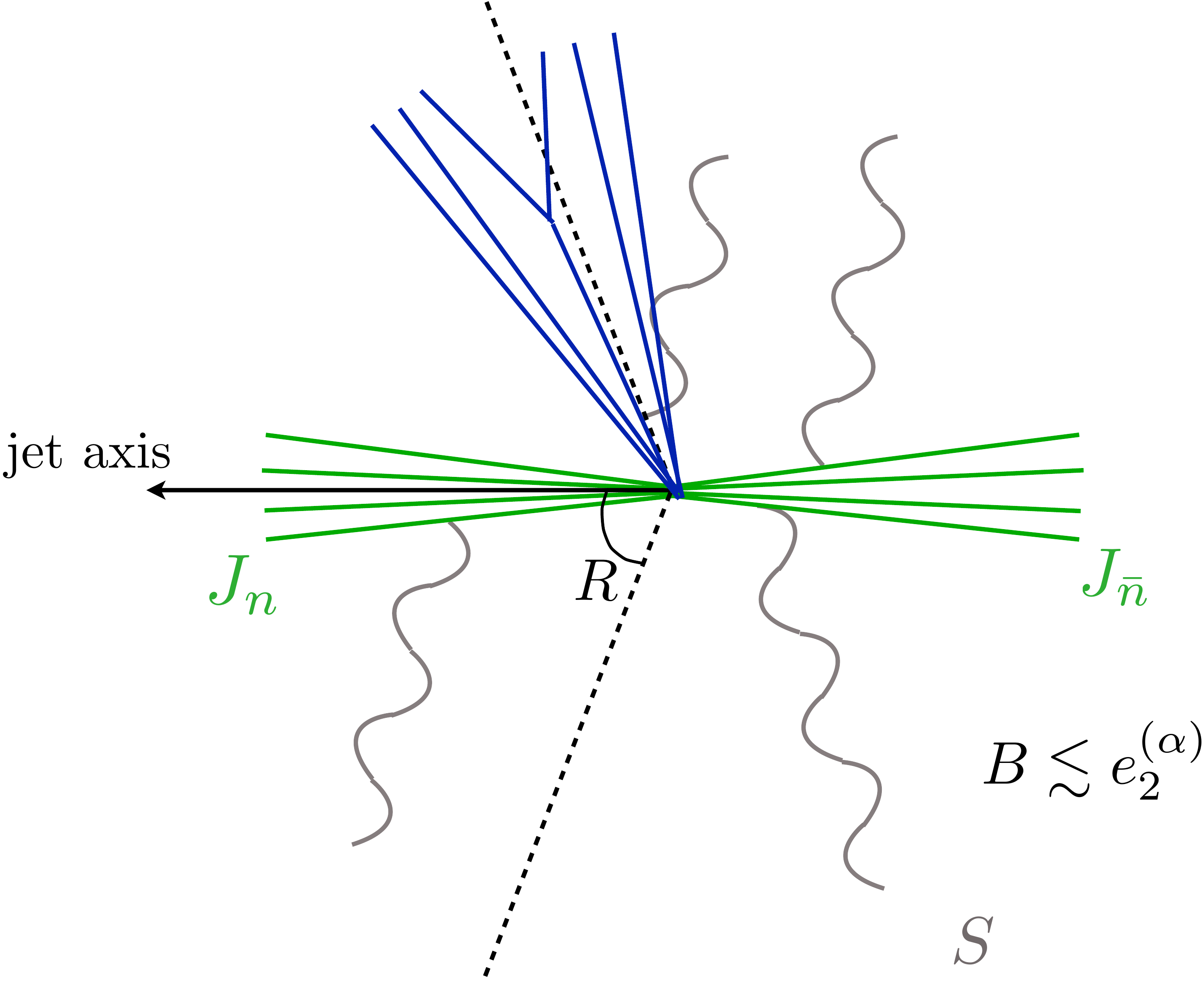}    
\end{center}
\caption{A soft jet at the edge of a jet. The dotted lines correspond to the jet boundary.
}
\label{fig:in_and_out}
\end{figure}

The specific power counting I adopt to isolate the one soft jet region of phase space at the edge of the jet is:
\begin{align}
\outj&\sim Q\ecf{2}{\alpha}\sim Q\ecf{2}{\beta}\\
\label{eq:subjet_region}\ecf{3}{\alpha}&\ll\Big(\ecf{2}{\alpha}\Big)^3\\
\label{eq:boundary_condition}(\Delta\sjtheta)^\alpha&\sim \frac{\ecf{3}{\alpha}}{(\ecf{2}{\alpha})^2}\ll 1
\end{align}
$\Delta\sjtheta=R-\sjtheta$ is the difference between the jet radius $R$ and the angle of the soft jet axis $\hat{n}_{sj}$ to the fat jet axis $\hat{n}$, denoted $\sjtheta$. This necessitates the soft jet to collinearly split into and out of the fat jet, pictorially represented in \Fig{fig:in_and_out}. Specifically, this implies the collinear modes of the soft jet in the soft jet factorization theorem of \cite{Larkoski:2015zka}, and the boundary soft modes, cannot be factorized, since the two obey the same scalings:
\begin{align}
p_{sj} &\sim E_J\, \ecfLa \left ( \left (  \frac{\ecfres}{  \left (\ecfLa\right)^2 }  \right)^{2/\alpha},1,\left (  \frac{\ecfres}{  \left (\ecfLa\right )^2 }  \right)^{1/\alpha}    \right )_{\sja\sjabar}   \,,\\
\label{eq:boundary_soft_scaling}
p_{bs} &\sim E_J\frac{  \ecfres   }{ \ecfLa   \left(\Delta \sjtheta\right)^\alpha }  \left ( \left(\Delta  \sjtheta\right)^2,1,\Delta  \sjtheta \right )_{\sja\sjabar}   \,, 
\end{align}
where I have adopted the light cone coordinates of the soft subjet, $(\sja\cdot p,\sjabar\cdot p, p_{\perp\sja})$, and $E_J\sim Q$ is the energy scale of the fat jet. 

\section{Factorization Theorem}
The factorization theorem takes the form:
\begin{align}\label{eq:collinear_splitting_fact}
\frac{d\sigma}{d\ecf{2}{\alpha}\,d\ecf{2}{\beta}\,d\ecf{3}{\alpha}d\outj}&=\sigma_0 H(Q^2)H_{n\bar{n}}^{sj}\Big(\ecf{2}{\alpha},\ecf{2}{\beta},\outj\Big)S_{n\nbar \sja}\Big(\ecf{3}{\alpha};R\Big)\nonumber\\
&\qquad\otimes\,\mathcal{E}_{\sja}\Big(\ecf{3}{\alpha};\outj;Q_{sj};R\Big)\otimes J_{n}(\ecf{3}{\alpha})\otimes J_{\nbar}(\outj)\otimes S_{n\nbar}\Big(\ecf{3}{\alpha};\outj;R\Big)
\end{align}
Where for conciseness, I have adopted the convention that any repeated argument is convolved with all other functions sharing that argument (excepting the jet radius $R$):
\begin{align}
F_1(x)\otimes F_2(x)...\otimes F_N(x)&=\int \prod_{i=1}^Ndx_i\delta\Big(x-\sum_{i=1}^Nx_i\Big)F_1(x_1)F_2(x_2)....F_N(x_N)
\end{align}
A new feature of the edge of jet factorization theorem is the edge of jet function $\mathcal{E}_{\sja}$. This is a novel jet function describing the fragmentation of partons into and out of the jet boundary. Though containing these fragmentation effects, the function is IR finite and calculable in perturbation theory, given the jet boundary cuts off the collinear divergence of the fragmentation.

I have also explicitly indicated the dependence on the soft subjet large momentum fraction $Q_{sj}$. Implicitly, this is a function of Q, $\ecf{2}{\alpha}$, and $\ecf{2}{\beta}$. The proof of this factorization theorem, like that for PDFs or fragmentation functions \cite{Bauer:2002nz}, takes place at the level of the cross-section or squared amplitude, not the amplitude itself, as in previous subjet factorizations. This is since the hard matching for $H_{n\bar{n}}^{sj}$ has both real and virtual corrections contributing to it. One constructs a complete basis of gauge invariant IR functions corresponding to all momentum regions. Since each sector has a unique leading power function according to the power counting, after including appropriate subtractions to remove overlaps, one is led to \eqref{eq:collinear_splitting_fact}. 

\subsection{Definitions of Factorized Functions}
I now give the operator definitions of the factorized low scale functions appearing in \Eq{eq:collinear_splitting_fact}. Each measurement function should be expanded in the power counting of the factorization theorem. The expanded versions can be found in \Ref{Larkoski:2015kga}.
\begin{itemize}
\item {\bf Tri-pole soft function:}
\begin{align}\label{eq:tripole_softs}
S_{n\nbar \sja}\Big(\ecf{3}{\alpha};R\Big)&=\int_{-i\infty+c}^{i\infty+c}d\ecflp{3}{\alpha}e^{\ecflp{3}{\alpha}\ecf{3}{\alpha}}\frac{\text{tr}\langle 0|T\{S_{\sja } S_{n} S_{\bar{n}}\}e^{-\ecflp{3}{\alpha}\Theta_{FJ}\ecfop{3}{\alpha}\big|_{S}}\bar{T}\{S_{\sja } S_{n} S_{\bar{n}}\} |0\rangle}{S_{n\nbar}\Big(\ecflp{3}{\alpha};R\Big)S_{\sja\sjabar}\Big(\ecflp{3}{\alpha};R\Big)}\\
S_{n\nbar}\Big(\ecflp{3}{\alpha};R\Big)&=\text{tr}\langle 0|T\{S_{n} S_{\bar{n}}\}e^{-\ecflp{3}{\alpha}\Theta_{FJ}\ecfop{3}{\alpha}\big|_{S}}\bar{T}\{S_{n} S_{\bar{n}}\} |0\rangle\\ 
S_{\sja\sjabar}\Big(\ecflp{3}{\alpha};R\Big)&=\text{tr}\langle 0|T\{S_{\sja} S_{\sjabar}\}e^{-\ecflp{3}{\alpha}\Theta_{FJ}\ecfop{3}{\alpha}\big|_{BS}}\bar{T}\{S_{\sja} S_{\sjabar}\} |0\rangle
\end{align}
To avoid convolutions between the necessary subtractions to the naive function, I give the definitions in laplace space. I have explicitly given the appropriate soft subtractions that must be performed on the tri-pole soft function. One removes both the boundary soft radiation associated with the soft subjet, and the soft radiation from the original $n$-$\nbar$ dipole. 
\item {\bf Edge-of-Jet Function:}
{\small\begin{align}
&\mathcal{E}_{\sja}\Big(\ecf{3}{\alpha};\outj;Q_{sj};R\Big)=\\
& \hspace{.25cm}
\frac{(2\pi)^3}{C_A}\text{tr}\langle 0|\mathcal{B}_{\perp_{\sja}}^{\mu}(0)\delta(\outj-\Theta_{\overline{FJ}}\mathbf{\outj})\delta(Q_{SJ}-\Theta_{FJ}\sjabar \cdot{\mathcal P})\delta^{(2)}(\vec{{\mathcal P}}_{\perp_{SJ}})\delta\Big(\ecf{3}{\alpha}-\Theta_{FJ}\ecfop{3}{\alpha}\big|_{SJ}\Big)\,\mathcal{B}_{\perp_{\sja}\mu}(0)|0\rangle \nonumber
\end{align}}
The jet region constraints $\Theta_{FJ}$ and $\Theta_{\overline{FJ}}$ restrictions project the state according to whether the momenta is inside the resolved fat jet, or the inclusive out-of-jet region, respectively, according to \Eq{eq:fat_jet_constraint}. The field operators are gauge invariant collinear gauge field strengths, whose definition can be found in \Ref{Larkoski:2014bxa}.
\item {\bf Dipole soft function:}
\begin{align}\label{eq:dipole_soft_function}
S_{n\nbar}\Big(\ecflp{3}{\alpha};\outj;R\Big)&=\text{tr}\langle 0|T\{S_{n} S_{\bar{n}}\}\delta\Big(\ecf{3}{\alpha}-\Theta_{FJ}\ecfop{3}{\alpha}\big|_{S}\Big)\delta(\outj-\Theta_{\overline{FJ}}\mathbf{\outj})\bar{T}\{S_{n} S_{\bar{n}}\} |0\rangle
\end{align}
Note that this function generates soft contributions to both $\outj$ and $\ecf{3}{\alpha}$.
\end{itemize}
The jet functions $J_n$ and $J_{\nbar}$ are standard for the given jet shape, see for instance \Refs{Hornig:2009vb,Ellis:2010rwa,Larkoski:2014uqa}.  

\section{Matching for Soft Jet Production}
\begin{figure}
\begin{center}
\includegraphics[width=4cm]{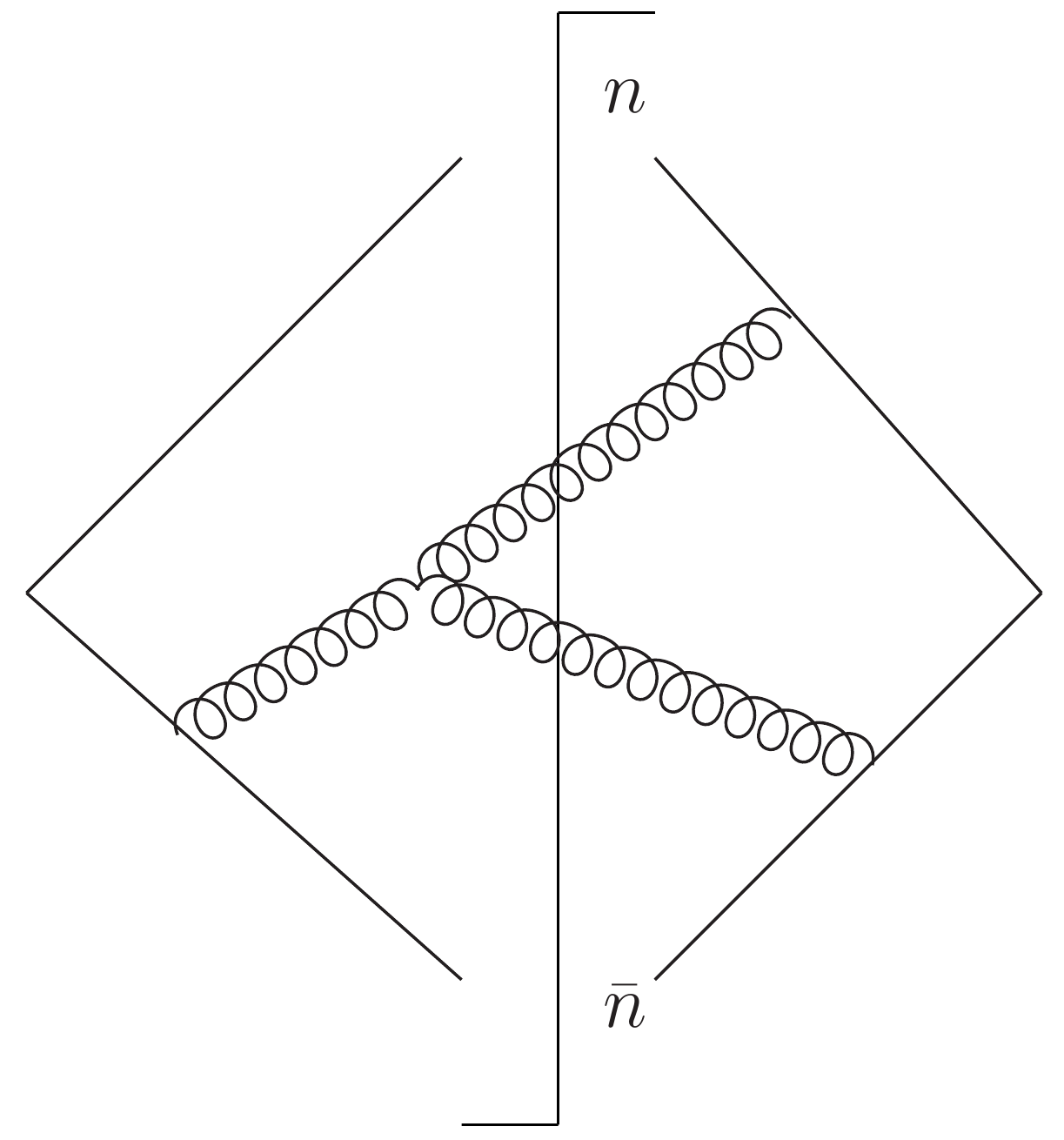}    
\end{center}
\caption{An example soft diagram that contributes to the soft jet production matching.
}
\label{fig:soft_diagram}
\end{figure}
Here I detail the one-loop matching procedure for the edge-of-jet factorization theorem\footnote{Throughout these calculations, I have made use of the Mathematica package HypExp \cite{Huber:2005yg, Huber:2007dx}.}. The soft emissions within the fat jet are strongly ordered with respect to the soft subjet, thus the calculations are identical to those of \cite{Larkoski:2015zka}. The virtual corrections can be extracted from \cite{Catani:2000pi}. However, the ``in-out'' configuration exhibits new terms, since $\outj\sim Q\ecf{2}{\alpha}\sim Q\ecf{2}{\beta}$. Therefore, the double real emission diagrams of full QCD are not expanded in a strongly ordered limit, but only in the soft limit compared to the hard scale $Q^2$. The integrand $S_{n\bar{n}}(p,q)$ for the double soft real emission is given in \cite{Catani:1999ss}. This corresponds to two soft gluon exchanges between the wilson lines in the $n$ and $\nbar$ directions, with \Fig{fig:soft_diagram} giving an example diagram. I can write the two real emission contribution to the fixed order cross-section as:
\begin{align}
d\sigma^{(2R)}&=N\int [d^dp]_+ [d^dq]_+\Big\{\Phi^{\text{in-in}}(\ecf{2}{\alpha},\ecf{2}{\beta},\ecf{3}{\alpha})S_{n\bar{n}}(p,q)\Big|_{S.O.}+\Phi^{\text{in-out}}(\ecf{2}{\alpha},\ecf{2}{\beta},\outj)S_{n\bar{n}}(p,q)\Big\}\\
 [d^dp]_+&=\frac{d^dp}{(2\pi)^{d-1}}\delta(p^2)\theta(p^0)
\end{align}
Where $\Phi^{\text{in-in}}$ and  $\Phi^{\text{in-out}}$ are the phase space constraints for when both emissions are inside the fat jet, and when one is inside the fat jet and the other out. ``S.O.'' denotes strong ordering. One emission will always be identified with the soft subjet, and to this end it is more convenient to trade the energy-energy correlation functions for the light-cone momentum fractions in the $n$-$\nbar$ coordinate system:
\begin{align}
 u&=n\cdot p_{sj},  &v&=\bar{n}\cdot p_{sj},\\
Q\ecf{2}{\alpha}&=(u+v)\Big(2\frac{u}{u+v}\Big)^{\frac{\alpha}{2}}, &Q\ecf{2}{\beta}&=(u+v)\Big(2\frac{u}{u+v}\Big)^{\frac{\beta}{2}} 
\end{align}
With an identified soft subjet, this is always possible up to power corrections, with a simple jacobian factor. 

The matching itself will be determined by the difference between the full theory matrix elements and the effective theory. That is, the one-loop matching to $H_{n\nbar}^{sj}$ is fixed by demanding:
\begin{align}
\frac{d\sigma}{du dv d\ecf{3}{\alpha} d\outj}=&H_{n\nbar}^{sj (T)}(u,v)\delta(\outj)\delta(\ecf{3}{\alpha})+H_{n\nbar}^{sj (T)}(u,v)\sum_{r}M_r^{(1)}\nonumber\\
&\qquad+H_{n\nbar}^{sj (1)}(u,v;\outj)\delta(\ecf{3}{\alpha})+O(\alpha_s^3)...
\end{align}
with the tree-level matching being:
\begin{align}\label{eq:tree-soft-jet}
H_{n\nbar}^{sj(T)}(u,v)&=\frac{\alpha_s C_F\mu^{2\epsilon}e^{\epsilon\gamma_E}}{\pi (uv)^{1+\epsilon}\Gamma(1-\epsilon)}=\frac{\alpha_s C_F}{2^{1+\epsilon}\pi\Gamma(1-\epsilon)}\Bigg(\frac{n\cdot \nbar}{n\cdot p_{sj}\,p_{sj}\cdot \nbar}\Bigg)^{1+\epsilon}
\end{align}
The sum $\sum_{r}M_r^{(1)}$ denotes the sum over one-loop matrix elements given by the infra-red functions of the factorization theorem \eqref{eq:collinear_splitting_fact}. The one-loop hard matching I split into three contributions, in-in, in-out, and in-virtual:
\begin{align}
H_{n\nbar}^{sj (1)}(u,v;\outj)&=H_{n\nbar}^{sj (1)}(u,v)\delta(\outj)\Big|_{\text{in-in}}+H_{n\nbar}^{sj (1)}(u,v;\outj)\Big|_{\text{in-out}}+H_{n\nbar}^{sj (1)}(u,v)\delta(\outj)\Big|_{\text{in-virt}}
\end{align}
One emission must always be in the fat jet, since the soft subjet has been observed. 
\subsection{In-In}
Restricting to the in-in configuration, in the strongly ordered expansion, the two real emission contribution is exactly reproduced by the factorization theorem contributions, so:
\begin{align}
H_{n\nbar}^{sj (1)}(u,v)\Big|_{\text{in-in}}&=0
\end{align}
This reflects the fact that the matching can be performed at the level of the amplitude, using only the IR finite terms of soft currents \cite{Larkoski:2015zka}.
\subsection{In-Virtual}
The real-virtual contribution is non-trivial. The infra-red divergent contributions are canceled by the virtual loops of the effective theory matrix elements, so that this contribution is completely determined by the finite parts of the soft-virtual contributions of the full theory. These are equivalent to the one-loop soft current, giving:
\begin{align}
H_{n\nbar}^{sj (1)}(u,v;\mu)\Big|_{\text{in-virt}}&=\frac{\alpha_s^2 C_F C_A}{\pi^2}(u\,v)^{-1}\Bigg(-\frac{1}{4}\text{ln}^2\Big(\frac{4\mu^2}{uv}\Big)+\frac{5}{24}\pi^2\Bigg)
\end{align}

\subsection{In-Out}
Here the full theory contribution is not fully reproduced by the effective theory, so that the difference between the effective theory and full gives a non-trivial contribution. I split the integrand into abelian and non-abelian components. The abelian contribution will be reproduced by the tree-level soft jet matching \eqref{eq:tree-soft-jet} and the out-of-fat-jet one-loop contribution from the dipole soft function \eqref{eq:dipole_soft_function}:
\begin{align}
d\sigma^{\text{in-out, ab.}}&=\Big(H_{n\bar{n}}^{sj (T)}(u,v)\Big)\Big(S_{n\bar{n}}^{(1)}(\outj)\Big|_{\text{out}}\Big)
\end{align}
The non-abelian piece has a more subtle matching contribution. I write:
\begin{align}\label{eq:in-out_matching_eqn}
d\sigma^{\text{in-out, n.ab.}}&=\Big(H_{n\bar{n}}^{sj (T)}(u,v)\Big)\Big({\mathcal E}_{\sja}^{(1)}(u+v;\outj;R)\Big|_{\text{in-out}}\Big)+H_{n\bar{n}}^{sj (1)}(u,v;\outj)\Big|_{\text{in-out}}\,.
\end{align}
Where I note that the soft jet large momentum fraction is given by $u+v=2Q_{sj}$. Next I explicitly calculate the full theory. This is simply the integral:
\begin{align}
d\sigma^{\text{in-out, n.ab.}}&=N\int [d^dp]_+ [d^dq]_+ \theta\Big(\frac{n\cdot q}{\bar{n}\cdot q}-\text{tan}^2\frac{R}{2}\Big)\delta\Big(\outj-\bar{n}\cdot q\Big)\nonumber\\
&\qquad\theta\Big(\text{tan}^2\frac{R}{2}-\frac{n\cdot p}{\nbar\cdot p}\Big)\delta(u-n\cdot p)\delta(v-\nbar\cdot p) S_{n\bar{n}}(p,q)\\
N=G^4C_F C_A(\mu e^{\gamma_E} 4\pi)^\epsilon
\end{align}
We are using the $\overline{MS}$ renormalization scheme, and $\gamma_E$ is Euler's gamma number. The integral measure can be written in terms of the relative angle between the transverse momenta components of $p$ and $q$, the other angle being trivial. Using the on-shell conditions to fix the magnitude of the transverse momentum integrals in terms of the light cone fractions, I then have for integral measure:
\begin{align}
&\int [d^dp]_+ [d^dq]_+\theta\Big(\frac{n\cdot q}{\bar{n}\cdot q}-\text{tan}^2\frac{R}{2}\Big)\delta\Big(\outj-\bar{n}\cdot q\Big)\theta\Big(\text{tan}^2\frac{R}{2}-\frac{n\cdot p}{\nbar\cdot p}\Big)\delta(u-n\cdot p)\delta(v-\nbar\cdot p)\nonumber\\
&\qquad=\Omega_{\epsilon}(u v)^{-\epsilon}\int_0^{\infty}\frac{dx}{x^\epsilon}\int_0^{\infty}\frac{dy}{y^{\epsilon}}\theta\Big(\frac{x}{y}-\text{tan}^2\frac{R}{2}\Big)\int_0^{\pi}d\theta\,\text{sin}^{-2\epsilon}\theta\delta\Big(\outj-y\Big)\,,\\
&\Omega_{\epsilon}=\frac{2^{-4+2\epsilon}\pi^{-\frac{9}{2}}}{\Gamma(1-\epsilon)\Gamma(\frac{1}{2}-\epsilon)}\,.
\end{align}
I have labeled the out-of-jet emission's light-cone momentum fractions as $x=n\cdot q$, and $y=\nbar\cdot q$. I now perform a further change of variables:
\begin{align}
x\rightarrow u x & & y\rightarrow v x y
\end{align}
Then with the further variable change, $x\rightarrow x(vy)^{-1}$, I can evaluate the $x,y$, and $\phi$ integrals as a series in $\epsilon$. Further, it is expedient to divide the in-out contribution into two pieces, isolating the collinear splitting scale $1-\frac{u}{v}\text{cot}^2\frac{R}{2}$ that will be removed by the effective theory subtractions.  The final result is, dropping all terms that vanish as $\epsilon\rightarrow 0$:
\begin{align}
d\sigma^{\text{in-out, n.ab.}}&=d\sigma^{\text{in-out, n.ab.}}_{\text{split}}+d\sigma^{\text{in-out, n.ab.}}_{\text{no-split}}\\ \label{eq:FT_In_Out_split}
d\sigma^{\text{in-out, n.ab.}}_{\text{split}}&=H_{n\nbar}^{sj(T)}(u,v)\frac{\alpha_s C_A}{\pi \outj}\Big(\frac{\mu^2 e^{\gamma_E}}{\outj^2}\frac{u}{v}\Big)^{\epsilon}\Bigg\{P_{}(v,\outj)\text{ln}\Big(1-\frac{u}{v}\text{cot}^2\frac{R}{2}\Big)\nonumber\\
&-\text{ln}\Big(1-\frac{u}{v}\text{cot}^2\frac{R}{2}\Big)+ \epsilon\text{ln}\Big(1-\frac{u}{v}\text{cot}^2\frac{R}{2}\Big)\text{ln}\Big(\frac{v}{u}\text{tan}^2\frac{R}{2}-1\Big)\Bigg\}+O\Big(\epsilon\Big)\\ \label{eq:FT_In_Out_no_split}
d\sigma^{\text{in-out, n.ab.}}_{\text{no-split}}&=H_{n\nbar}^{sj(T)}(u,v)\frac{\alpha_s C_A}{\pi \outj}\Big(\frac{\mu^2 e^{\gamma_E}}{\outj^2}\frac{u}{v}\Big)^{\epsilon}\Bigg\{P_{}(v,\outj)\text{ln}\Big(\frac{\outj}{\outj+u\,\text{cot}^2\frac{R}{2}}\Big)\nonumber\\
&+\frac{ u v \outj(v-\outj)\text{cot}^2\frac{R}{2}}{(v+\outj)^3(\outj+u \text{cot}^2\frac{R}{2})}\Big(1-2\frac{T_fn_f}{C_A}\Big)-\frac{\outj}{2(v+\outj)}\text{ln}\Big(\frac{\outj}{\outj+u\,\text{cot}^2\frac{R}{2}}\Big)\Bigg\}+O\Big(\epsilon\Big)
\end{align}
I have explicitly factored out the all-orders in $\epsilon$ tree-level hard matching. Where $P(v,\outj)$ is given by:
\begin{align}
P(v,\outj)=-\frac{(v^2+v\outj+\outj^2)^2}{(v+\outj)^4}-\frac{T_fn_f}{C_A}\frac{v\outj(v^2+\outj^2)}{(v+\outj)^4}+1
\end{align}
Note that $P(v,\outj)$ is finite as $\outj\rightarrow 0$, and the no-split contribution has no soft divergence as $\outj\rightarrow 0$. It is the gluon's $1\rightarrow 2$ splitting function with soft support removed. One can explicitly check taking the laplace/fourier transforms of $u,\outj$, and integrating over the allowed $v$, these terms with $R=\frac{\pi}{2}$ are sufficient to reproduce the hemisphere NGLs of \eqref{eq:FO_hemi_NGLS}, and both the split and no-split terms contribute. 

To make the divergent behavior of this contribution to the cross-section transparent, I take the laplace transform of out of jet variable $\outj$ in \Eq{eq:FT_In_Out_split}, whose conjugate variable I denote by $\outjlp$. Then I can simply expand in $\epsilon$, without making use of plus distributions\footnote{I factor out the all-orders in $\epsilon$ tree level matching. The higher orders in $\epsilon$ do not contribute to the matching, even though they naively give $O(\epsilon^0)$ terms when multiplied against the $\epsilon^{-1}$ terms of the one-loop correction. However, since all $\epsilon$ divergences cancel when I sum over sectors, and the higher order terms in epsilon from the tree-level matching multiple all sectors identically, these spurious finite terms cancel out in the total cross-section.}.
\begin{align} \label{eq:FT_In_Out_split_II}
d\tilde\sigma^{\text{in-out, n.ab.}}_{\text{split}}&=\int_0^{\infty}d\outj \,e^{-\outj\outjlp}d\sigma^{\text{in-out, n.ab.}}_{\text{split}}\\
&=H_{n\nbar}^{sj(T)}(u,v)\frac{\alpha_s C_A}{\pi}\text{ln}\Big(1-\frac{u}{v}\text{cot}^2\frac{R}{2}\Big)\Bigg\{\frac{1}{2\epsilon}+\frac{1}{2}\text{ln}\Bigg(\frac{e^{2\gamma_E}\mu^2\outjlp^2}{\text{tan}^2\frac{R}{2}-\frac{u}{v}}\Bigg)+\tilde P\Big(v\outjlp\Big)\Bigg\}+O(\epsilon)\\
\tilde P\Big(v\outjlp\Big)&=\int_0^{\infty}\frac{d\outj}{\outj}\, e^{-\outj\outjlp}P(v,\outj)
\end{align} 
One can work out the exact analytic expression for $\tilde P$ in terms of incomplete gamma functions and exponentials, but the form is relatively unenlightening, other than it is a function of the product $v \outjlp$ only. To complete the matching calculation, I need the edge-of-jet jet function, which is taken up in the next section.

\section{Edge-of-Jet Function}

I start this section with a general discussion of the phase space for the edge of jet function. Then I focus on the calculation for the contribution relevant for the matching and the resummation of NGLs, the in-out configuration. The in-in configuration is relegated to an appendix.

\subsection{Tree Level and One-Loop Phase Space for Edge-of-Jet Function}
The tree level Edge-of-Jet function is calculated to be:
\begin{align}
{\mathcal E}_{\sja}^{(T)}\Big(\ecf{3}{\alpha};\outj;Q_{sj};R\Big)&=\delta(\ecf{3}{\alpha})\delta(\outj)(2\pi)^3\int [d^4p]_+\delta^{(2)}(p_\perp)\delta(Q_{sj}-\sjabar\cdot p)\nonumber\\
&=\delta(\ecf{3}{\alpha})\delta(\outj)\,.
\end{align}
The transverse momentum constraint localizes the tree-level result to align with the observed soft jet axis, so that no contribution is given to $\outj$ at tree-level. Moving to one loop, I have:
\begin{align}
{\mathcal E}_{\sja}^{(1)}\Big(\ecf{3}{\alpha};\outj;Q_{sj};R\Big)&=g^2C_A\Big(2\pi\Big)^{d-1}\int [d^dk_1]_+ [d^dk_2]_+ \Big(\Phi^{\text{in-in}}(\ecf{3}{\alpha},Q_{sj},R,k_1,k_2)\delta(\outj)\nonumber\\
&\qquad+\Phi^{\text{in-out}}(\outj,Q_{sj},R,k_1,k_2)\delta(\ecf{3}{\alpha})\Big)\frac{ P_{g}(\sjabar\cdot k_1,\sjabar\cdot k_2)}{k_1\cdot k_2}
\end{align}
where the splitting function for an off-shell gluon into two partons with momenta $k_{1}, k_{2}$ is given as:
\begin{align}
\frac{ P_{g}(\sjabar\cdot k_1,\sjabar\cdot k_2)}{k_1\cdot k_2}&=-2\frac{\big((\sjabar\cdot k_1)^2+\sjabar\cdot k_1\sjabar\cdot k_2+(\sjabar\cdot k_2)^2\big)^2}{\sjabar\cdot k_1\,\sjabar\cdot k_2(\sjabar\cdot k_1+\sjabar\cdot k_2)^2\,k_1\cdot k_2}\nonumber\\
&\qquad\qquad+ \frac{T_fn_f}{C_A}\frac{2\sjabar\cdot k_1\sjabar\cdot k_2-(1-\epsilon)(\sjabar\cdot k_1+\sjabar\cdot k_2)^2}{(1-\epsilon)(\sjabar\cdot k_1+\sjabar\cdot k_2)^2k_1\cdot k_2}
\end{align} 
The two phase space restrictions break down as:{\small
\begin{align}
\Phi^{\text{in-in}}(\ecf{3}{\alpha},Q_{sj},R,k_1,k_2)&=\delta(Q_{sj}-\sjabar\cdot k_1-\sjabar\cdot k_2)\delta\Bigg(\ecf{3}{\alpha}-N\frac{\sjabar\cdot k_1}{Q}\frac{\sjabar\cdot k_1}{Q}\Big(\frac{2 k_1\cdot k_2}{\sjabar\cdot k_1\sjabar\cdot k_2}\Big)^{\frac{\alpha}{2}}\Bigg)\nonumber\\
&\qquad\delta^{(d-2)}(k_{1\perp}+k_{2\perp})\theta\Big(\Delta\theta_{sj}+\frac{4n_{\perp}\cdot k_{1\perp}}{\sjabar\cdot k_1 (\nbar\cdot\sja)^2}\Big)\theta\Big(\Delta\theta_{sj}+\frac{4n_{\perp}\cdot k_{2\perp}}{\sjabar\cdot k_2 (\nbar\cdot\sja)^2}\Big)\\
\Phi^{\text{in-out}}(\outj,Q_{sj},R,k_1,k_2)&=\delta(Q_{sj}-\sjabar\cdot k_1)\delta\Big(\outj-\frac{\bar{n}\cdot\sja}{2}\sjabar\cdot k_2 \Big)\delta^{(d-2)}(k_{1\perp}+k_{2\perp})\nonumber\\
&\qquad\theta\Big(\Delta\theta_{sj}+\frac{4n_{\perp}\cdot k_{1\perp}}{\sjabar\cdot k_1 (\nbar\cdot\sja)^2}\Big)\theta\Big(\Delta\theta_{sj}-\frac{4n_{\perp}\cdot k_{2\perp}}{\sjabar\cdot k_2 (\nbar\cdot\sja)^2}\Big)
\end{align}}
All transverse projections $\perp$ are with respect to the local soft jet light-cone coordinates defined by $\sja$ and $\sjabar$. I have introduced the angle to the jet boundary $\Delta\theta_{sj}$:
\begin{align}
\Delta\theta_{sj}=\text{tan}^2\frac{R}{2}-\frac{n\cdot\sja}{\nbar\cdot\sja}
\end{align}
The $\theta$-function constraints come from expanding the fat jet constraint \Eq{eq:fat_jet_constraint}, in the coordinate frame of the soft subjet when it is pressed against the boundary:
\begin{align}
\theta\Big(\text{tan}^2\frac{R}{2}-\frac{n\cdot k}{\nbar\cdot k}\Big)&=\theta\Big(\Delta\theta_{sj}+\frac{4n_{\perp}\cdot k_{2\perp}}{\sjabar\cdot k_2 (\nbar\cdot\sja)^2}\Big)+...
\end{align}
To avoid expansions in plus-distributions, I will henceforth work with the laplace transform of $\ecf{3}{\alpha}$, $\ecflp{3}{\alpha}$. 

\subsection{In-Out Contribution}
The bare contribution is given by:
\begin{align}
{\mathcal E}_{\sja}^{(1)}\Big(\outj;u,v;R\Big)\Bigg|_{\text{in-out}}&=\frac{\alpha_sC_A}{4^{1-\epsilon}\epsilon}\frac{\text{sec }\pi\epsilon}{\pi^{1/2}\Gamma(\frac{1}{2}-\epsilon)}\frac{v}{(v+\outj)^2}\Big(\frac{\mu^2 e^{\gamma_E}}{\outj^2(\text{tan}^2\frac{R}{2}-\frac{u}{v})}\frac{u}{v}\Big)^{\epsilon}P_{g}\Big(v,\outj\Big)
\end{align}
Again, I want to take the laplace transform to extract the divergences. To do so I first isolate the soft contribution, by writing:
\begin{align}
P_{g}\Big(v,\outj\Big)&=2\frac{(v+\outj)^2}{v\outj}\Big(-1+P(v,\outj)\Big)
\end{align}
So:
\begin{align}
{\mathcal E}_{\sja}^{(1)}\Big(\outj;u,v;R\Big)\Bigg|_{\text{in-out}}&=-\frac{\alpha_sC_A}{2^{1-2\epsilon}\epsilon}\frac{\text{sec }\pi\epsilon}{\pi^{1/2}\Gamma(\frac{1}{2}-\epsilon)}\frac{1}{\outj}\Big(\frac{\mu^2 e^{\gamma_E}}{\outj^2(\text{tan}^2\frac{R}{2}-\frac{u}{v})}\frac{u}{v}\Big)^{\epsilon}\nonumber\\
&\qquad+\frac{\alpha_sC_A}{2^{1-2\epsilon}\epsilon}\frac{\text{sec }\pi\epsilon}{\pi^{1/2}\Gamma(\frac{1}{2}-\epsilon)}\frac{1}{\outj}\Big(\frac{\mu^2 e^{\gamma_E}}{\outj^2(\text{tan}^2\frac{R}{2}-\frac{u}{v})}\frac{u}{v}\Big)^{\epsilon}P(v,\outj)\,.
\end{align}
Now I take the laplace transform, and expand in $\epsilon$, to achieve:
\begin{align}\label{eq:Edge_Of_Jet_div_in_out}
{\mathcal E}_{\sja}^{(1)}\Big(\outjlp;u,v;R\Big)\Big|_{\text{in-out}}^{div}&=-\frac{C_A\alpha_s}{\pi}\Bigg\{\frac{1}{4\epsilon^2}+\frac{1}{2\epsilon}\text{ln}\Big(\frac{e^{\gamma_E}\mu\outjlp}{\text{tan}^2\frac{R}{2}-\frac{u}{v}}\sqrt{\frac{u}{v}}\Big)+\frac{1}{2\epsilon}\tilde P(v\outjlp)\Bigg\}\\
\label{eq:Edge_Of_Jet_fin_in_out}{\mathcal E}_{\sja}^{(1)}\Big(\outjlp;u,v;R\Big)\Big|_{\text{in-out}}^{fin}&=-\frac{\alpha_sC_A}{\pi}\Bigg\{\tilde P(v\outjlp)\text{ln}\Bigg(\frac{\mu}{v(\text{tan}^2\frac{R}{2}-\frac{u}{v})}\sqrt{\frac{u}{v}}\Bigg)-\tilde P^{\text{ln}}(v\outjlp)\nonumber\\
&\qquad+\frac{1}{2}\text{ln}^2\Big(\frac{e^{\gamma_E}\mu\outjlp}{\text{tan}^2\frac{R}{2}-\frac{u}{v}}\sqrt{\frac{u}{v}}\Big)+\frac{7\pi^2}{48}+I_{n_f}(v,\outjlp)\Bigg\}\\
\tilde P^{\text{ln}}(v\outjlp)&=\int_0^{\infty}\frac{d\outj}{\outj}\,e^{-\outj\outjlp}P(v,\outj)\text{ln}\frac{\outj}{v}\\
I_{n_f}(v\outjlp)&=\frac{T_fn_f}{C_A}\int_0^{\infty}\frac{d\outj}{\outj}\,e^{-\outj\outjlp}\frac{v^2\outj^2}{(v+\outj)^4}
\end{align}
The finite terms in momentum space are:
\begin{align}
\label{eq:Edge_Of_Jet_fin_in_out_momentum_space}{\mathcal E}_{\sja}^{(1)}\Big(\outj;u,v;R\Big)\Big|_{\text{in-out}}^{fin}&=-\frac{\alpha_sC_A}{\pi}\Bigg\{\frac{P(v,\outj)}{\outj}\text{ln}\Bigg(\frac{\mu}{\outj(\text{tan}^2\frac{R}{2}-\frac{u}{v})}\sqrt{\frac{u}{v}}\Bigg)\nonumber\\
&\qquad+{\mathcal L}^1\Big(\frac{\mu}{\text{tan}^2\frac{R}{2}-\frac{u}{v}}\sqrt{\frac{u}{v}},\outj\Big)+\frac{\pi^2}{16}\delta(\outj)+\frac{T_fn_f}{C_A}\frac{v^2\outj}{(v+\outj)^4}\Bigg\}
\end{align}
The in-out contribution to the edge-of-jet function has several noteworthy features. Aside from containing double logarithmic divergences, it also features a single logarithmic divergence corresponding to the non-soft terms of DGLAP splitting function. At higher orders one can expect departures from DGLAP evolution. The appearance of DGLAP is not suprising, since essentially at this order in perturbation theory, the measurements imposed give the probability for a jet close to the boundary to fragment a parton out of the fat jet boundary. The IR divergence associated with this fragmentation is cut off by the angular distance of the soft jet to the fat jet boundary, so though the function will have an evolution similar to a fragmentation function, it is IR-finite and calculable in perturbation theory. 

It is natural that the double logarithmic divergences, here being related to the soft limit of the splitting functions, are controlled by an anomalous dimension proportional to the cusp to all orders in perturbation theory \cite{Polyakov:1980ca,Brandt:1981kf,Korchemsky:1987wg,Korchemsky:1992xv}. 

\subsection{Edge-of-Jet Function Matching}\label{sec:matching_edge_of_jet_function}
The edge-of-jet function admits an OPE onto the boundary soft function and a standard jet function, of the form:
\begin{align}\label{matching_edge_of_jet_function}
{\mathcal E}_{\sja}\Big(\ecf{3}{\alpha};\outj;Q_{sj};R\Big)&=C_{\sja}(\outj,R)S_{\sja\sjabar}(\ecf{3}{\alpha},R)J_{\sja}(\ecf{3}{\alpha},Q_{sj})+O\Bigg(Q\ecf{3}{\alpha}\Big(1-\frac{u}{v}\text{cot}^2\frac{R}{2}\Big)\Bigg)
\end{align}
The tree-level matching is:
\begin{align}
C_{\sja}^{(T)}(\outj,R)&=\delta(\outj)\,.
\end{align}
Since the boundary softs and standard jet function calculations reproduce eqns. \eqref{eq:edge_of_jet_in_in_stnd_start} to \eqref{eq:edge_of_jet_in_in_bsoft_fin}, the divergent and finite contributions in laplace space to the matching coefficient to one-loop accuracy is then given by \Eq{eq:Edge_Of_Jet_div_in_out} and \Eq{eq:Edge_Of_Jet_fin_in_out}:
\begin{align}
C_{\sja}^{(1)}(\outjlp,R)\Big|^{\text{div}}&={\mathcal E}_{\sja}^{(1)}\Big(\outjlp;u,v;R\Big)\Big|_{\text{in-out}}^{\text{div}}\\
C_{\sja}^{(1)}(\outjlp,R)\Big|^{\text{fin}}&={\mathcal E}_{\sja}^{(1)}\Big(\outjlp;u,v;R\Big)\Big|_{\text{in-out}}^{\text{fin}}\label{edge_of_jet_matching_JET_FUNCTION}
\end{align}
The renomalization group equation satisfied by this renormalized matching coefficient is to one loop:
\begin{align}
\mu\frac{d}{d\mu}\text{ln}C_{\sja}(\outjlp,R)&=-\frac{\alpha_s(\mu)C_A}{\pi}\Bigg\{\text{ln}\Big(\frac{e^{\gamma_E}\mu\outjlp}{\text{tan}^2\frac{R}{2}-\frac{u}{v}}\sqrt{\frac{u}{v}}\Big)+\tilde P(v\outjlp)\Bigg\}
\end{align}
The anomalous dimensions of the other functions in \Eq{matching_edge_of_jet_function} are given in \Ref{Larkoski:2015zka}, and one can construct the full RG equation of ${\mathcal E}_{\sja}$ from those results.

\section{Edge-of-Jet Matching}
I now have all the ingredients to finish the matching in \Eq{eq:in-out_matching_eqn}. I use \eqref{eq:FT_In_Out_split_II}, \eqref{eq:Edge_Of_Jet_div_in_out}, and \eqref{eq:Edge_Of_Jet_fin_in_out} to compute:
\begin{align}
H^{sj(1)}_{n\nbar}(u,v,\outjlp)\Big|_{\text{in-out}}^{\text{div}}&=H_{n\nbar}^{sj(T)}(u,v)\frac{\alpha_sC_A}{\pi}\Bigg\{\frac{1}{4\epsilon^2}+\frac{1}{2\epsilon}\text{ln}\Big(\frac{\mu\outjlp e^{\gamma_E}}{\text{tan}^2\frac{R}{2}}\sqrt{\frac{u}{v}}\Big)+\frac{1}{2\epsilon}\tilde P(v\outjlp)\Bigg\}\\
\label{eq:in_out_matching_coef}H^{sj(1)}_{n\nbar}(u,v,\outjlp)\Big|_{\text{in-out}}^{\text{fin}}&=H_{n\nbar}^{sj(T)}(u,v)\frac{\alpha_sC_A}{\pi}\Bigg\{+\frac{1}{2}\text{ln}^2\Big(\frac{\mu\outjlp e^{\gamma_E}}{\text{tan}^2\frac{R}{2}}\sqrt{\frac{u}{v}}\Big)\nonumber\\
&\qquad-\text{ln}\Big(\frac{u}{v\text{tan}^2\frac{R}{2}}\Big)\text{ln}\Big(1-\frac{u}{v\text{tan}^2\frac{R}{2}}\Big)+\frac{7\pi^2}{48}\nonumber\\
&\qquad+\tilde P(v\outjlp)\text{ln}\Bigg(\frac{\mu}{v\text{tan}^2\frac{R}{2}}\sqrt{\frac{u}{v}}\Bigg)-\tilde P^{\text{ln}}(v\outjlp)-I_{n_f}(v\outjlp)\nonumber\\
&\qquad+\text{no split}\Bigg\}\,.
\end{align}
I also give the finite contribution in momentum space in terms of plus distributions. This allows us to show the structure of the ``no-split'' terms more transparently:
\begin{align}
\label{eq:in_out_matching_coef_momentum}H^{sj(1)}_{n\nbar}(u,v,\outj)\Big|_{\text{in-out}}^{\text{fin}}&=H_{n\nbar}^{sj(T)}(u,v)\frac{\alpha_sC_A}{\pi}\Bigg\{\delta(\outj)\Big(-\text{ln}\Big(\frac{u}{v\text{tan}^2\frac{R}{2}}\Big)\text{ln}\Big(1-\frac{u}{v\text{tan}^2\frac{R}{2}}\Big)+\frac{\pi^2}{16}\Big)\nonumber\\
&+{\mathcal L}^{1}\Big(\mu\,\text{cot}^2\frac{R}{2}\sqrt{\frac{u}{v}},\outj\Big)+\frac{1}{\outj}P(v,\outj)\text{ln}\Bigg(\Big(\frac{\mu}{\outj\text{tan}^2\frac{R}{2}}\sqrt{\frac{u}{v}}\Big)\Big(\frac{\outj}{\outj+u\text{ cot}^2\frac{R}{2}}\Big)\Bigg)\nonumber\\
&-\frac{\outj}{2(v+\outj)}\text{ln}\Big(\frac{\outj}{\outj+u\,\text{cot}^2\frac{R}{2}}\Big)\nonumber\\
&+\frac{ u v \outj(v-\outj)\text{cot}^2\frac{R}{2}}{(v+\outj)^3(\outj+u \text{cot}^2\frac{R}{2})}\Big(1-2\frac{T_fn_f}{C_A}\Big)+T_fn_f\frac{v^2\outj}{(v+\outj)^4}\Bigg\}
\end{align}
The plus distributions are defined in \App{Plus_Distros}. Importantly, this contribution to the matching coefficient is manifestly finite as $u\rightarrow v\text{tan}^2\frac{R}{2}$. That is, all sensitivity to the IR scales associated with the collinear splitting at the edge of the jet have been removed.
\section{One Soft Jet Contribution to NGLs with Collinear Effects}
\subsection{Fixed Order Non-Global Logarithms}
For reference, I quote the fixed order result at $\alpha_s^2$ for hemisphere NGLs, see \Refs{ Hornig:2011iu,Kelley:2011ng}. Dividing out the global contribution to the cross-section, the hemisphere NGL contribution for an initial dipole in the fundamental representation is:
\begin{align}\label{eq:FO_hemi_NGLS}
\sigma^{NGL}\Big(\frac{u^c}{\outj^c}\Big)\Big|_{u^c\gg \outj^c}&=\frac{\alpha_s^2}{\pi^2}C_F\Bigg(-\frac{\pi^2}{12}C_A\text{ln}^2\frac{u^c}{\outj^c}+\text{ln}\frac{u^c}{\outj^c}\Big(C_A\frac{11\pi^2-3-18\zeta_3}{36}+T_Rn_f\frac{6-4\pi^2}{36}\Big)\Bigg)
\end{align}
One can check by explicit calculation that the result for $N=4$ SYM is simply the leading terms in transcendentality. The leading NGL at $\alpha_s^2$ for non-hemispherical jets were obtained for a variety of algorithms in \Ref{Hornig:2011tg}, and both leading and subleading were obtained for cone algorithms, similar to the ones considered in this paper, in \Ref{Kelley:2011aa}. 

It is important to note that the $\zeta_3$ subleading log is directly tied to the collinear double logarithms of the edge of jet function found in \eqref{eq:Edge_Of_Jet_fin_in_out}. In particularly, it arises in the full theory calculation of the NGLs integrating over the double logarithm at order $\epsilon$ in the second line of \Eq{eq:FT_In_Out_split}. This fact will play an important role in the collinear evolution of the BMS kernel later.

\subsection{Dressing Gluons}
To elucidate the collinear splitting contributions to the NGLs, I following the procedure outlined in \cite{Larkoski:2015zka}. I re-associate functions in the factorization theorem \eqref{eq:collinear_splitting_fact} to obtain anomalous dimensions that allows me to resum the NGLs for a single soft jet emitted off of the $n$-$\bar{n}$ dipole. I define:
\begin{align}\label{eq:defining_W_func}
W_{n\bar{n}}(p_{sj};\outj;R)&=\lim_{\ecflp{3}{\alpha}\rightarrow 0} H^{sj}_{n\nbar}\Big(p_{sj};\outj;R\Big){\mathcal E}_{\sja}\Big(\ecflp{3}{\alpha}\Big)S_{\sja\sjabar}\Big(\ecflp{3}{\alpha};R\Big)S_{n\nbar\sja}\Big(\ecflp{3}{\alpha};R\Big)\\
\label{eq:defining_G_func}G_{n\nbar\sja}(\outj;R)&=C_{\sja}(\outjlp,R)
\end{align}
These functions differ from those used in the original soft jet factorization to resum NGLs. Because of the modified power counting, both $W$ and $G$ can depend on the out of jet scale, since formally I have assumed no hierarchy. The RG equations satisfied by these functions are given by:
\begin{align}
\label{eq:RG_W_func}\mu\frac{d}{d\mu}\text{ln}W_{n\bar{n}}(p_{sj};\outjlp;R)&=\frac{\alpha_s(\mu)C_A}{\pi}\Bigg\{\text{ln}\Big(\frac{\mu\outjlp e^{\gamma_E}}{\big(\text{tan}^2\frac{R}{2}-\frac{u}{v}\big)}\sqrt{\frac{u}{v}}\Big)+\tilde P(v\outjlp)\Bigg\}\\
\label{eq:RG_G_func}\mu\frac{d}{d\mu}\text{ln}G_{n\nbar\sja}(\outjlp;R)&=-\frac{\alpha_s(\mu)C_A}{\pi}\Bigg\{\text{ln}\Big(\frac{\mu\outjlp e^{\gamma_E}}{\big(\text{tan}^2\frac{R}{2}-\frac{u}{v}\big)}\sqrt{\frac{u}{v}}\Big)+\tilde P(v\outjlp)\Bigg\}
\end{align}
I can see how the RG-improvement of the $WG$ product resums the collinear splittings:
\begin{multline}
W_{n\bar{n}}(p_{sj};\outjlp;R;\mu)G_{n\nbar\sja}(\outjlp;R;\mu)=W_{n\bar{n}}(p_{sj};\outjlp;R,\mu)G_{n\nbar\sja}(\outjlp;R;\mu_i)\tilde U(\mu,\mu_i,;\outjlp)\\
\end{multline}
It is convenient to split the resummation factor into a double and single logarithmic contributions as:
\begin{align}
\tilde U(\mu,\mu_i;\outjlp)&=\tilde U_{\text{ln}^2}(\mu,\mu_i;\outjlp)\tilde U_{\text{ln}}(\mu,\mu_i;\outjlp)\\
\tilde U_{\text{ln}^2}(\mu,\mu_i;\outjlp)&=\text{Exp}\Bigg[-\int_{\mu_i}^{\mu}\frac{d\mu'}{\mu'}\frac{\alpha_s(\mu')C_A}{\pi}\text{ln}\Big(\frac{\mu}{\mu_i}\Big)\nonumber\\ 
&\qquad-\text{ln}\Big(\frac{\mu_i\outjlp e^{\gamma_E}}{\big(\text{tan}^2\frac{R}{2}-\frac{u}{v}\big)}\sqrt{\frac{u}{v}}\Big)\int_{\mu_i}^{\mu}\frac{d\mu'}{\mu'}\frac{\alpha_s(\mu')C_A}{\pi}\Big)\Bigg]\\
\tilde U_{\text{ln}}(\mu,\mu_i;\outjlp)&=\text{Exp}\Bigg[-\tilde P(v\outjlp)\int_{\mu_i}^{\mu}\frac{d\mu'}{\mu'}\frac{\alpha_s(\mu')C_A}{\pi}\Bigg]
\end{align}
I can then write the cumulant momentum space resummed distribution as a convolution between the single and double logarithmic functions:
\begin{align}
U^c(\mu,\mu_i;\outj)&=\int_0^{\infty}d\outj'\,U_{\text{ln}^2}^c(\mu,\mu_i;\outj^c-\outj')U_{\text{ln}}(\mu,\mu_i;\outj')
\end{align}
I cannot obtain $U_{\text{ln}}(\mu,\mu_i;\outj)$ in analytic form, but I can solve for $U_{\text{ln}^2}^c(\mu,\mu_i;\outj^c)$:
\begin{align}
\label{eq:dl_cumulant_evo}U_{\text{ln}^2}^c(\mu,\mu_i;\outj^c)&=\theta(\outj^c)\Big(\frac{\outj^c}{\mu}\Big)^{-\omega(\mu,\mu_i)}\frac{\text{Exp}\Big[-K(\mu,\mu_i)\Big]}{\Gamma(1-\omega(\mu,\mu_i))}\\
\omega(\mu,\mu_i)&=\int_{\mu_i}^{\mu}\frac{d\mu'}{\mu'}\frac{\alpha_s(\mu')C_A}{\pi}\nonumber\\
&\sim\frac{\alpha_s C_A}{\pi}\text{ln}\frac{\mu}{\mu_i}\\
K(\mu,\mu_i)&=\text{ln}\Big(\frac{\mu_ie^{\gamma_E}}{\mu\big(\text{tan}^2\frac{R}{2}-\frac{u}{v}\big)}\sqrt{\frac{u}{v}}\Big)\int_{\mu_i}^{\mu}\frac{d\mu'}{\mu'}\frac{\alpha_s(\mu')C_A}{\pi}\nonumber\\
&\sim\frac{\alpha_s C_A}{\pi}\text{ln}\Big(\frac{\mu}{\mu_i}\Big)\text{ln}\Big(\frac{\mu_ie^{\gamma_E}}{\mu\big(\text{tan}^2\frac{R}{2}-\frac{u}{v}\big)}\sqrt{\frac{u}{v}}\Big)
\end{align}
The single soft jet contribution to the cumulative non-global logarithms then takes in resummed form:
\begin{align}\label{eq:collinear_splitting_ngls}
\sigma^{\text{NGL}}\Big(\frac{u^c}{\outj^c}\Big)=\int_{\outj^c}^{u^c}du\int_u^\infty dv\int_{u-\outj^c}^{\outj^c}d\outj\Big\{&W_{n\nbar}(u,v;\outj;\mu)\otimes G(u,v;\outj;\mu_i)\nonumber\\
&\otimes U_{\text{ln}^2}^c(\mu,\mu_i;\outj)\otimes U_{\text{ln}}(\mu,\mu_i;\outj)-\text{c-bin}\Big\}
\end{align}
The limits of integration are determine as follows. The variable $v$  is just integrated over all allowed angles of the fat jet, while $u, \outj$ are bounded as follows:
\begin{align}
\outj^c>\outj & & u^c>u \\
\outj+u>\outj^c & &u>\outj^c
\end{align}
The conditions of the first line are just the definition of the cumulative values of the measurement. The second line is more important. It states that the parent gluon that splits in and out of the jet is above the IR scale of the problem. Thus the edge of jet resummation can be applied without interfering with the global and IR divergences.  Below the IR scale, and the gluon contributes to the global divergence, and no resummation should be applied. This also guarantees that all IR divergences cancel between the $in-in$, $in-out$, and $out-out$ regions for the hemisphere soft function \cite{Hornig:2011iu}.

To see whether the dressed gluon will reproduce the $\alpha_s^2$ NGLs, I must expand each of these functions to first nontrivial order, and evaluate the integrals. Note that the collinear bin is the expansion in the limit that $u\ll v,\outj$ before the integrals are evaluated \cite{Manohar:2006nz}. This removes any overlap with a collinear subjet factorization. To this order (mixed leading-logarithmic prime, since we are incorporating matching contributions, and $\alpha_s$ running, but ignoring the two-loop cusp contribution), the collinear-bin insures that the limit of \Eq{eq:defining_W_func} is well defined. Any part of the one-loop matrix elements contributing to the resolution variable $\ecf{3}{\alpha}$ will cancel against the same contribution with the collinear-bin expansion performed, up to corrections that are beyond the logarithmic order I am working. 
\subsubsection{Matrix Element Contributions to NGLs}
Due to the collinear bin, the only contribution to the NGLs that can arise from the matrix elements of the effective theory are those of either the in-out contribution to the soft jet production coefficient in \Eq{eq:in_out_matching_coef}, or the edge-of-jet matching coefficient of \Eq{edge_of_jet_matching_JET_FUNCTION}. Thus we have to order $\alpha_s$ in each of the low scale functions:
\begin{align}
W_{n\nbar}(u,v;\outj;\mu)G_{n\nbar\sja}(\outj;R;\mu_i)&=H_{n\bar{n}}^{sj(T)}(u,v)\delta(\outj)+H_{n\bar{n}}^{sj(1)}(u,v,\outj;\mu)\Big|_{\text{in-out}}\nonumber\\
&\qquad+H_{n\bar{n}}^{sj(T)}(u,v)C_{\sja}^{(1)}(\outj;R;\mu_i)+O(\alpha_s^2)
\end{align}
Examining the matrix element in \Eq{eq:Edge_Of_Jet_fin_in_out_momentum_space}, one can choose the low scale to be:
\begin{align}
\label{eq:low_scale_choice_canonical}
\mu_i\sim \outj^c\Big(\text{tan}^2\frac{R}{2}-\frac{u}{v}\Big)\sqrt{\frac{v}{u}}
\end{align}
This will minimize all logs in the edge-of-jet function that can contribute to the NGLs. After the collinear bin, no terms from the edge-of-jet function will contribute to the $\alpha_s^2$ NLGs. Examining the soft jet production matching coefficient, taking:
\begin{align}
\label{eq:high_scale_choice_canonical}
\mu\sim \outj^c\sqrt{\frac{v}{u}}\,\text{tan}^2\frac{R}{2}\,,
\end{align}
will minimize the logarithms in \Eq{eq:in_out_matching_coef_momentum}. However, with this choice, there are still terms left over that can contribute to the NGLs even after the collinear bin:
\begin{align}\label{eq:matrix_element_for_ngls}
H_{n\bar{n}}^{sj(1)}(u,v,\outj;\mu)\Big|_{\text{in-out}}-\text{c-bin}=\frac{\alpha_s^2}{\pi^2}&\frac{C_FC_A}{u v}\Bigg(+\frac{1}{\outj}P(v,\outj)\text{ln}\Big(\frac{\outj}{\outj+u\text{ cot}^2\frac{R}{2}}\Big)\nonumber\\
&-\delta(\outj)\text{ln}\Big(\frac{u}{v\text{tan}^2\frac{R}{2}}\Big)\text{ln}\Big(1-\frac{u}{v\text{tan}^2\frac{R}{2}}\Big)\nonumber\\
&-\frac{\outj}{2(v+\outj)}\text{ln}\Big(\frac{\outj}{\outj+u\,\text{cot}^2\frac{R}{2}}\Big)\nonumber\\
&+\frac{ u v \outj(v-\outj)\text{cot}^2\frac{R}{2}}{(v+\outj)^3(\outj+u \text{cot}^2\frac{R}{2})}\Big(1-2\frac{T_fn_f}{C_A}\Big)\Bigg)
\end{align}
In the hemisphere case, setting $R=\pi/2$, I evaluate the resulting integrals to get:
\begin{align}
&\int_{\outj^c}^{u^c}\frac{du}{u}\int_u^\infty \frac{dv}{v}\int_{u-\outj^c}^{\outj^c}\frac{d\outj}{\outj}\,\text{ln}\Big(\frac{\outj}{u+\outj}\Big)P\Big(v,\outj\Big)=
\label{eq:NGL_Other_Split_Terms}\Big(\frac{22\pi^2+9+6\text{ ln}2+132\text{ ln}^22}{144}
\nonumber\\ &\qquad\qquad\qquad-\frac{n_fT_R}{C_A}\frac{4\pi^2+9+6\text{ ln}2+24\text{ ln}^22}{72}\Big)\text{ln}\frac{u^c}{\outj^c}+...\\
\label{eq:NGL_Zeta3_Terms}&\int_{\outj^c}^{u^c}\frac{du}{u}\int_u^\infty \frac{dv}{v}\int_{u-\outj^c}^{\outj^c}\frac{d\outj}{\outj}\,\frac{-\outj}{2(\outj +v)}\text{ln}\Big(\frac{\outj}{u+\outj}\Big)=-\frac{\zeta_3}{2}\text{ln}\frac{u^c}{\outj^c}+...\\
\label{eq:NGL_Rational_Terms}&\int_{\outj^c}^{u^c}\frac{du}{u}\int_u^\infty \frac{dv}{v}\int_{u-\outj^c}^{\outj^c}\frac{d\outj}{\outj}\,\frac{v-\outj}{2(u+\outj)(\outj +v)^3}\Big(1-2\frac{n_fT_R}{C_A}\Big)=-\frac{3}{16}\Big(1-2\frac{n_fT_R}{C_A}\Big)\text{ln}\frac{u^c}{\outj^c}+...
\end{align}
Where we have dropped terms that are finite as $\frac{u^c}{\outj^c}\rightarrow \infty$. Note that the large logarithms can be found simply by setting:
\begin{align}\label{eq:dropping_subleading_terms}
\int_{u-\outj^c}^{\outj^c}d\outj\rightarrow -\int_0^ud\outj+...
\end{align}
Then the finite terms are automatically dropped. This naturally coincides with an ordering of emissions $u>\outj$ found in the evolution equation approach.

\subsubsection{Evolution Factors Contribution to NGLs}
Next I must expand the cumulant resummed distribution with scale choices given by \Eqs{eq:low_scale_choice_canonical}{eq:high_scale_choice_canonical}. I also take the collinear-bin and subtract. I find the following integrals to evaluate:
\begin{align}
&\text{Expanding }U_{\text{ln}^2}^c(\mu,\mu_i;\outj_c)-U_{\text{ln}^2}^c(\mu,\mu_i;u-\outj_c)-\text{c-bin}:\nonumber\\
\label{eq:NGL_DoubleL_Terms}&\int_{\outj^c}^{u^c}\frac{du}{u}\int_u^\infty \frac{dv}{v}\text{ln}\Big(1-\frac{u}{v}\Big)\text{ln}\Big(\frac{u-\outj^c}{\outj^c}\Big)=-\frac{\pi^2}{12}\text{ln}^2\frac{u^c}{\outj^c}+...\\
\nonumber\\
&\text{Expanding }U_{\text{ln}}(\mu,\mu_i;\outj)-\text{c-bin}:\nonumber\\
\label{eq:NGL_Split_Terms}&\int_{\outj^c}^{u^c}\frac{du}{u}\int_u^\infty \frac{dv}{v}\int_{u-\outj^c}^{\outj^c}\frac{d\outj}{\outj}\,\text{ln}\Big(1-\frac{u}{v}\Big)P\Big(v,\outj\Big)=\Big(\frac{11\pi^2+3-3\text{ ln}2-66\text{ ln}^22}{72}
\nonumber\\ &\qquad\qquad\qquad\qquad\qquad\qquad-\frac{n_fT_R}{C_A}\frac{4\pi^2+6-6\text{ ln}2-24\text{ ln}^22}{72}\Big)\text{ln}\frac{u^c}{\outj^c}+...
\end{align}
Again we have dropped terms that are finite as $\frac{u^c}{\outj^c}\rightarrow \infty$. Upon summing Eqs. \eqref{eq:NGL_Other_Split_Terms} through \eqref{eq:NGL_Split_Terms}, and restoring the color factors and coupling, we reproduce \Eq{eq:FO_hemi_NGLS}.

\subsection{Alternative Resummation Schemes}
One can attempt to exponentiate more terms than is done with the canonical choice of factorization scale \Eq{eq:high_scale_choice_canonical}. In particular, one is tempted to set:
\begin{align}
\mu\sim \outj^c+u\,
\end{align}
This choice formally exponentiates all terms that can be connected to the splitting function, moving \eqref{eq:NGL_Other_Split_Terms} into the single log resummation factor. However, the terms connected with the rational terms \eqref{eq:NGL_Rational_Terms} would still not be exponentiated. Given that the logarithm associated with these terms is effectively a ratio of soft energy scales, $\outj$ to $\outj+u$, a more systematic route to the resummation of these terms would be to calculate and resum the subleading power corrections to the soft jet factorization of \cite{Larkoski:2015zka}, along the lines of \cite{Larkoski:2014bxa,Laenen:2008gt,Laenen:2008ux,Laenen:2010uz,Bonocore:2015esa}.

\section{NLO BMS}
The comparison to the recently derived NLO BMS equation found in \cite{Caron-Huot:2015bja} is straightforward\footnote{I thank Simon Caron-Huot for correspondence about the NLO procedure for BMS, as well as providing notes on the calculations.}. Within this approach, one seeks to define an equation of motion for the reduced density matrix that results from integrating out the hard emissions. One does this time step by time step, where each previous step the mode that resulted from a decay into the IR now becomes a hard eikonal current, freely radiating. This results in a Markovian process at leading lo, and departures from leading log can be understood as corrections to the Markovian picture. The  ``time'' evolution of the reduced density matrix is governed by a hamiltonian, and one can simply examine the action of the two loop hamiltonian generating the evolution on the initial dipole, perform the requisite IR averaging, and drop any terms that occur at order greater than $\alpha_s^2$. 

For simplicity, we restrict to $N=4$ SYM, where the hamiltonian is simplest. If
\begin{align}
K&=\frac{\alpha_s}{4\pi}K^{(1)}+\frac{\alpha_s^2}{16\pi^2}K^{(2)}+...
\end{align} 
is the hamiltonian generating the soft evolution, I have for this hamiltonian acting on an initial dipole in $N=4$ SYM:
\begin{align}\label{eq:Caron-Huot_NLO_kernel}
K^{(2)}\text{Tr}[U_1U_2^\dagger]&=2\int\frac{d^2\Omega_0}{4\pi}\frac{d^2\Omega_{0'}}{4\pi}K_{12;00'}^{(2)N=4,l}\Big(2U_{0}^{aa'}U_{0'}^{aa'}-U_{0'}^{aa'}U_{0'}^{aa'}-U_{0}^{aa'}U_{0}^{aa'}\Big)\nonumber\\
&\qquad\times\Big(\text{Tr}\Big[[T^a,T^b]U_1T^{a'}T^{b'}U_2^\dagger\Big]+\text{Tr}\Big[T^{a}T^{b}U_1[T^{a'},T^{b'}]U_2^\dagger\Big]\Big)\nonumber\\
&+\frac{4\pi^2 C_A}{3}\int\frac{d^2\Omega_0}{4\pi}\frac{\alpha_{12}}{\alpha_{10}\alpha_{02}}\Big(\text{Tr}[T^{a'}U_1T^aU^\dagger_2]U_0^{aa'}-C_A\text{Tr}[U_1U_2^\dagger]\Big)
\end{align}
Where I have:
\begin{align}
\alpha_{a b}&=-\frac{\beta_a\cdot \beta_b}{2}\\
\beta_a&=(1,\hat{b}_a),\qquad 1=\hat{b}_a^2\\
K_{n\nbar;00'}^{(2)N=4,l}&=\frac{\alpha_{n\nbar}}{\alpha_{0n}\alpha_{00'}\alpha_{0'\nbar}}\Bigg(2\text{ln}\Big(\frac{\alpha_{n\nbar}\alpha_{00'}}{\alpha_{0\nbar}\alpha_{0'n}}\Big)+\Big[1+\frac{\alpha_{n\nbar}\alpha_{00'}}{\alpha_{0n}\alpha_{0'\nbar}-\alpha_{0'n}\alpha_{0\nbar}}\Big]\text{ln}\Big(\frac{\alpha_{0n}\alpha_{0'\nbar}}{\alpha_{0'n}\alpha_{0\nbar}}\Big)\Bigg)
\end{align}
That is, all integrals are over null rays on the celestial sphere. I have gone ahead and set the initial dipole to be the back to back jets. The first term corresponds to multiple soft emissions generating new wilson lines $U_0, U_{0'}$, while the second term corresponds to the cusp contribution. The IR averaging is accomplished by:
\begin{align}\label{eq:Caron-Huot_strong_ordering}
U_i^{a b}\rightarrow \delta^{ab}\Bigg(\theta\Big(\frac{p_i^z}{p_i^0}\Big)+\theta\Big(\text{cos}R-\frac{p_i^z}{p_i^0}\Big)\Bigg)
\end{align}
I focus on the multiple soft emission term, since this will naturally produce the single logarithmic contributions. Noting that one can drop the small cone about the other hemisphere, the integrations become with the IR averaging constraints:
\begin{align}
2\int\frac{d^2\Omega_0}{4\pi}\frac{d^2\Omega_{0'}}{4\pi}\rightarrow_{\text{IR ave.}}\int_0^{1}dx\int_0^{1}dy\int_{0}^{2\pi}\frac{d\phi}{4\pi}\Bigg(\theta\Big(\frac{1}{2}-x\Big)\theta\Big(y-\frac{1}{2}\Big)+\theta\Big(x-\frac{1}{2}\Big)\theta\Big(\frac{1}{2}-y\Big)\Bigg)
\end{align} 
Here, $x,y$ are related to the polar angles of the soft emissions. The NGL distribution is generated by acting Exp$[-L_{NGL}\hat{K}]$ on the initial hard configuration, so expanding the exponential and using the results in \cite{Caron-Huot:2015bja} for the kernel, we get:
{\small \begin{align}\label{eq:Simon_LO_ave_Nis4}
-\frac{\alpha_s^2}{16\pi^2C_A}\Bigg\langle K^{(2)N=4}\text{Tr}[U_1U_2^\dagger]\Bigg\rangle_{IR}\text{ln}\frac{u^c}{\outj^c} &=\frac{\alpha_s^2C_AC_F}{2\pi^2}\zeta_3\text{ln}\frac{u^c}{\outj^c}
\end{align}}%
When calculating with a hamiltonian germane to QCD, one obtains almost \Eq{eq:FO_hemi_NGLS}, except for the $\zeta_3$ term above, which has the wrong sign. In the full BMS equation, there is also the contribution coming from iteration of the leading order hamiltonian, however, this simply generates the leading log NGL at two loops. Then to get the $\zeta_3$ logarithm correct, one must supplement the LO IR averaging procedure with an NLO averaging correction. However, we will see that collinearly improving the BMS equation to contain the leading double collinear logs results in a contribution that when added to the above, gives the correct $\zeta_3$, allowing us to forgo the NLO averaging procedure\footnote{One must be careful here since there is a great deal of scheme dependence as to where one puts the corrections for the NLLs, between the $K^{(1)}, K^{(2)}$ evolution hamiltonians, versus the IR calculation of the observable after evolution. Only the sum of the three are scheme invariant, and expected to give the full answer.}. This is not unexpected: under the assumption that the LO and NLO BMS kernels fully capture all soft coherence in the IR evolution, the only remaining corrections in the IR that can be obtained are incoherent emissions off of the wilson lines. 

\section{Collinear Improvement of the BMS Equation}\label{sec:collinear_BMS}
It is beyond the scope of this paper to firmly establish how one might collinearly-improve the NLO BMS equation, however, noting the correspondence between the soft jet expansion in \cite{Larkoski:2015zka} and the leading order BMS equation, one can derive an expression that will hopefully capture all double collinear logs, at least in the large $N_c$ limit\footnote{Since the collinear evolution is derived from color singlet objets, jet functions, this implies the collinear evolution does not care about the planar limit. However, the exact matching scale can be sensitive to the directions of the soft wilson lines the soft jet is entangled with.}\footnote{The above NLO equation is for full color evolution, however, for checking the two loop NGLs, this is immaterial.}. The purely non-global version of BMS is given as:
\begin{align}
\label{eq:BMS}
\mu\frac{d}{d\mu} g_{a b}&=\frac{\alpha_s C_A}{\pi}\int_{FJ}\frac{d\Omega_{j}}{4\pi}W_{ab}(j)\Big\{U_{abj}\,g_{aj}g_{jb}-g_{ab}\Big\}\,,\\
W_{ab}(j)&=\frac{a\cdot b}{a\cdot n_j\,n_j\cdot b}\,.
\end{align}
Where $a,b$, and $n_j$ are null vectors. The integration is over the angular region of the fat jet. The RG equations of the soft jet factorization could be used to generate the $U_{abj}$ factor:
\begin{align}
\label{eq:BMS_anom_dim_ab}
\mu\frac{d}{d\mu}\text{ln}U_{abj}&=-\frac{\alpha_sC_A}{\pi}\int_{\overline{FJ}}\frac{d\Omega_{i}}{4\pi}\Big\{W_{aj}(i)+W_{jb}(i)-W_{ab}(i)\Big\}&=-\frac{\alpha_s C_A}{\pi}S_{ab}(j)\,.
\end{align}
Where the integration is over the angular region of out of jet region. From \Ref{Schwartz:2014wha}, the function $S_{ab}(j)$ always contains the term ln$\Big(1-\frac{u}{v}\Big)$. Indeed, examining \Eqs{eq:RG_W_func}{eq:RG_G_func} and noting that for the case of $a=n, b=\nbar$:
\begin{align}
\label{eq:BMS_anom_dim_nnbar}
\int_{\overline{FJ}}\frac{d\Omega_{i}}{4\pi}\Big\{W_{nj}(i)+W_{jn}(i)-W_{n\nbar}(i)\Big\}&=-\text{ln}\Big(1-\frac{u}{v}\Big)\\
\frac{u}{v}=\frac{n\cdot n_j}{\nbar\cdot n_j}
\end{align}
I will show how to collinearly improve the BMS equation, assuming that the effective theories correctly calculate the resummed production rate of soft jets modulo global effects. This resummation of the collinear double logs maintains the conformal invariance of the BMS equation, up to the boundary condition of the jet radius. I construct the resummed weight in the BMS equation through a sequence of effective theories under the assumption:
\begin{align}
\label{eq:collinear_regime}\outj^c(1-\frac{u}{v})\ll \outj^c\,,\\
\label{eq:standard_BMS_regime}\outj^c \ll \mu\,.
\end{align}
I evolve between the scales of \Eq{eq:collinear_regime} using the evolution the logarithmic terms of \Eq{eq:RG_G_func}, then I evolve between the scales of \Eq{eq:standard_BMS_regime} using \Eqs{eq:BMS_anom_dim_nnbar}{eq:BMS_anom_dim_nnbar}. This sequence of evolutions can be codified into a limit of the factorization theorem in \cite{Larkoski:2015zka}, where one essentially takes the threshold limit of the collinear splittings at the jet boundary, which is given in \App{app:threshold_soft_jet}. Then we have for the resummation in the BMS equation:
\begin{align}\label{eq:collinear_improved_U_factor}
U_{n\nbar j}^{\text{c.i.}}&=\text{Exp}\Bigg[\frac{\alpha_s C_A}{\pi}\text{ln}\Big(1-\frac{u}{v}\Big)\text{ln}\Big(\frac{u}{\outj^c}\Big)-\frac{\alpha_s C_A}{2\pi}\text{ln}^2\Big(1-\frac{u}{v}\Big)\Bigg]
\end{align}
Then the collinearly improved BMS equation for the $n-\nbar$ dipole is:
\begin{align}
\mu\frac{d}{d\mu} g_{n \nbar}&=\frac{\alpha_s C_A}{\pi}\int_{FJ}\frac{d\omega_{j}}{4\pi}W_{n\nbar}(j)\Big\{U_{n \nbar j}^{\text{c.i.}}\,g_{nj}g_{j\nbar}-g_{n\nbar}\Big\}\,.
\end{align}
 Finally, one would like to collinearly improve the BMS equation for an arbitrary dipole with directions $a,b$,  derived in \App{app:threshold_soft_jet}:
\begin{align}\label{eq:collinear_improved_U_factor_arb_dipole}
U_{ab j}^{\text{c.i.}}&=\text{Exp}\Big[-\frac{\alpha_s C_A}{\pi}S_{ab}(j)\text{ln}\Big(\frac{\mu}{\outj^c}\Big)+\frac{\alpha_s C_A}{\pi}S_{ab}(j)\text{ln}\Big(1-\frac{u}{v}\Big)+\frac{\alpha_s C_A}{2\pi}S_{ab}^2(j)\Big]
\end{align} 
One can check from the explicit expressions for $S_{ab}(j)$ (see \App{Out_Of_Jet_Anom}) this scheme also exponentiates the same double collinear log. Expanding out this scheme to find the predicted logs, one finds it over-estimates the $\zeta_3$ NGL by a factor of two. Namely, expanding out the collinear improved BMS equation and integrating, we find the NGL contribution:
\begin{align}
\frac{\alpha_sC_A}{2\pi}\int_{\outj^c}^{u^c}\frac{du}{u}\int_u^{\infty}\frac{dv}{v}\Big(\frac{\alpha_s C_A}{\pi}\text{ln}\Big(1-\frac{u}{v}\Big)\text{ln}\Big(\frac{\mu}{\outj^c}\Big)-\frac{\alpha_s C_A}{2\pi}\text{ln}^2\Big(1-\frac{u}{v}\Big)\Big)\nonumber\\
=\frac{\alpha_s^2C_A^2}{2\pi^2}\Big(-\frac{\pi^2}{12}\text{ln}^2\frac{u^c}{\outj^c}-\zeta_3\text{ln}\frac{u^c}{\outj^c}\Big)
\end{align}
Correcting for the large $N_c$ limit by swapping $C_A^2\rightarrow 2C_F C_A$, and adding this to \Eq{eq:Simon_LO_ave_Nis4}, I find that the leading in transcendantality terms of \Eq{eq:FO_hemi_NGLS} (the $N=4$ SYM result) are reproduced. The specific to QCD contributions are already captured in the formalism of \cite{Caron-Huot:2015bja}. Thus the collinear improvement of the BMS equation removes the need to supplement the NLL resummation of NGLs with IR averaging corrections, that is, no large log still resides in the IR after evolution has taken place. The scheme dependence of the IR averaging has been completely shuffled into a redefinition of the BMS kernels, though there will now exist a scheme dependence between the collinear evolution, and the fixed order expression for the BMS kernels. The collinear resummation scales chosen here (the natural ones from the point of view of the factorization theorem) correspond to the ``Lorentz'' scheme evolution kernel $K^{(2)}$.

Here we have only used the leading singular pieces of the splitting function for the collinear improvement, however, other than having to solve a more complicated evolution equation, the full edge of jet function contains all the ingredients necessary to also capture these effects. Careful attention should be paid to the resulting changes in the $K^{(2)}$ kernel if further subleading collinear resummation is performed.

\section{Collinear Matching for Deep versus Edge Subjets, and Finding the Buffer Region}
I seek to argue that the collinear corrections to the BMS equation for jets deep in the fat jet are irrelavant, exactly so in a conformal theory like $N=4$ SYM. Every wilson line generate by the BMS evolution naturally comes with its own jet function, if for no other reason than to absorb and cancel collinear divergences. The question then is whether these jet functions at all have a natural IR scale beyond that of the fat jet's mass (which sets their energy scale), or are the``inclusive-unobserved'' jet functions of \Ref{Ellis:2010rwa}. Given no natural IR scale, any loop integral in these jet functions will be scaleless, and in dimension regularization, they will be zero. In the calculation of the BMS kernel, they will appear as at most collinear corrections that cancel divergences in ``hard'' virtual corrections of these kernels. In QCD, there will always be the IR scale of $\Lambda_{QCD}$, so though the collinear integrals will be scaleless in perturbation theory, running to non-perturbative scales is still necessary. Angles of subjets with respect to each other are always bounded by the jet radius, with no enhancing singularity, thus never give a meaningful scale. As subjets approach each other, they are simply combined, and collinear divergences cancel with no left over large logarithm. Then only close to the fat jet boundary does a new IR scale arises: the angle to this boundary that cuts off fragmentation into or out of the jet. This now gives the collinear corrections a non-trivial structure, even in a conformal theory, and one can conjecture that all scheme dependence of the IR averaging will arise from corrections having to do with edge of jet physics, since that is the only large scale left in the problem.  Thus with the appropriate collinear resummation of the BMS kernel, one could render the IR averaging procedure trivial, in the sense that it is always given by the leading order procedure, or the leading order procedure times a kinematically trivial series in $\alpha_s$. Moreover, these corrections are universally predicted from a single jet function calculation. From an effective theory viewpoint, this is pleasing, since one would like to obtain all large logs from evolution, with no surprises in the IR trace of the reduced density matrix.

\begin{figure}
\begin{center}
\includegraphics[scale=.75]{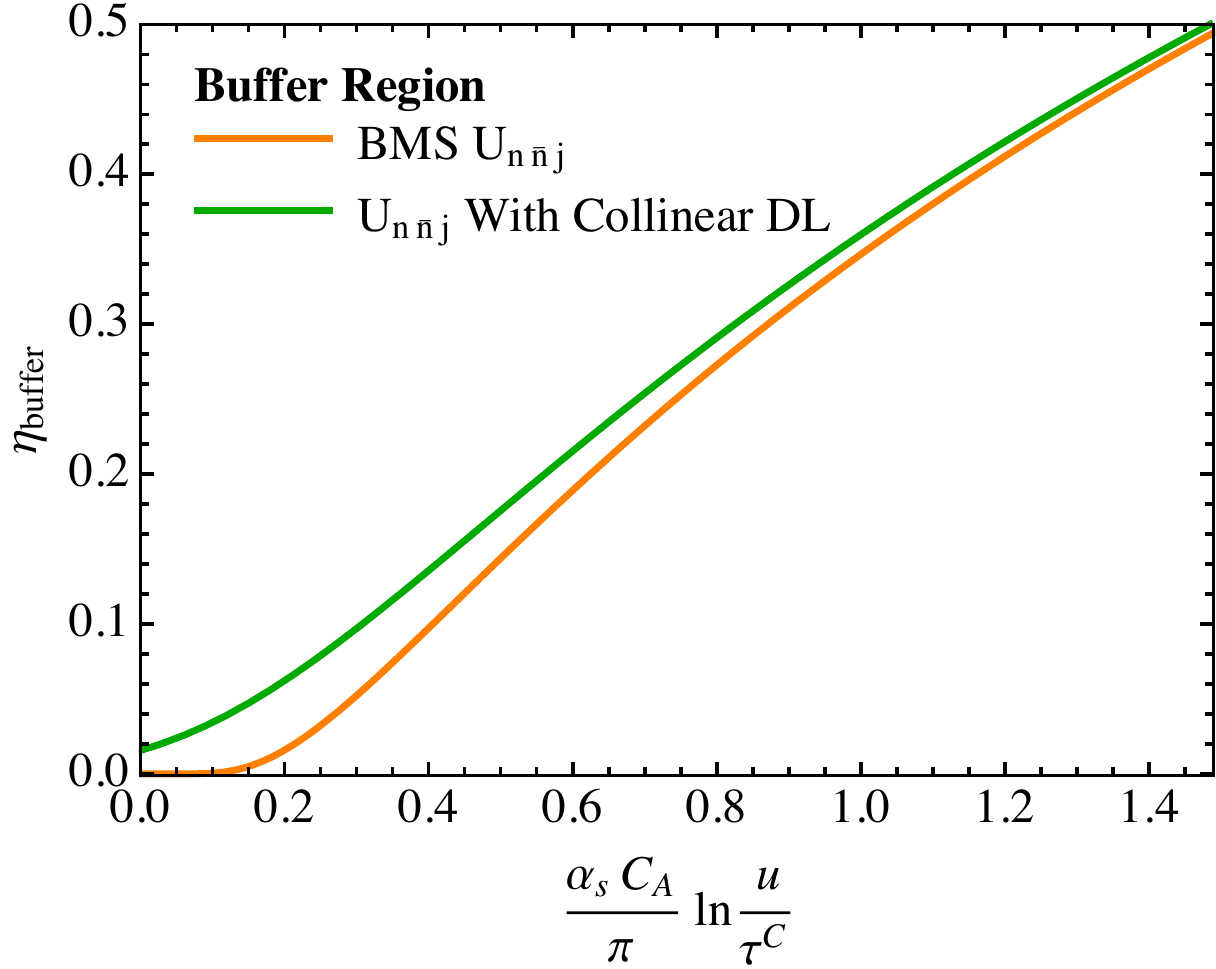}    
\end{center}
\caption{The rapidity of the soft jet at which the resummation factor in \Eq{eq:BMS} reaches half its maximum value. Plotted are both the original BMS resummation, and the collinear improvement for the $n$-$\nbar$ dipole. Note: $\frac{\alpha_s C_A}{\pi}\sim .11$.
}
\label{fig:Buffer}
\end{figure}

Finally, it is important to quantify how big the edge of the jet is. Other than the potential confinement scale, only the size of the NGL can matter. So we have the different angular regimes:
\begin{align}
1-\frac{u}{v}\ll \frac{\outj^c}{\mu} \text{ or }  1-\frac{u}{v}\gg \frac{\outj^c}{\mu}
\end{align}
In the first, the second term of \Eq{eq:collinear_improved_U_factor} dominates the exponent, while in the latter the standard term dominates. Thus the size of the non-global log effectively sets the size of the boundary layer containing the collinear theory. The larger the NGL, the less the collinear resummation contributes to the improved equation. This is just the statement that the leading logs are genuinely leading. However, at small to moderate values of the NGL, the collinear effects could be expected to be substantial. One can get a feel for this competition by plotting out the size of the buffer region. This is defined to be the rapidity of the soft jet at which the resummation factor $U$ in \Eq{eq:BMS} attains its half maximum, given the size of the NGL. This is plotted in \Fig{fig:Buffer}.

\section{The Lindblad Equation and the Expansion in Soft Jets}\label{sec:Lindblad}
These non-linear IR/RG evolution equations found in jet/small-x physics all fit a very general pattern (see also \cite{Nagy:2007ty,Waalewijn:2012sv,Dasgupta:2014yra}). Namely, since the underlying physics is that of a Markov process at leading log, the equations are examples of the Lindblad equation \Refs{KOSSAKOWSKI1972247,Lindblad:1975ef,Gorini:1976cm}.  These evolution equations for the reduced density matrix in the presence of an energetic enviroment decaying into the IR take a universal form, and in a field theory context have recently been discussed in \Refs{Burgess:2014eoa,Agon:2014uxa,Boyanovsky:2015xoa}. If one takes a generic unitary quantum system with some initial energetic state with density matrix $\hat{\rho}$, and begins to integrate out the energetic modes of the matrix, giving a reduced density matrix $\hat{\rho}_R$, then evolution in the ``time'' $t$ (or whatever observable the reduced density matrix is to be ordered in) satisfies:
\begin{align}
\partial_t \hat{\rho}_R&= [i\hat{H},\hat{\rho}_R]+\sum_{ij} \Gamma_{ij}(t)\Big\{2\hat{L}_{i}\hat{\rho}_R\hat{L}_{j}^{\dagger}-\hat{L}_{j}^{\dagger}\hat{L}_{i}\hat{\rho}_R-\hat{\rho}_R\hat{L}_{j}^{\dagger}\hat{L}_{i}\Big\}\,
\end{align}
if and only if the system is Markovian, and the evolution satisfies a semi-group composition rule while preserving the positivity of the reduced density matrix. $\hat{H}$ is a hermitian operator, and the evolution is non-unitary (the flow is diffusive and potentially irreversible, thus the \emph{semi-group} composition rule of the evolution.), and this is in contrast to traditional RG equations found in factorization theorems where one can flow in either direction through RG consistency. The $\Gamma_{ij}$ are the decay rates of the energetic enviroment into the IR, and $\hat{L}_i$ create/annihilate the asymptotic IR states. The sum over $i,j$ is over the kinematically allowed phase space for the decay, assuming a mode can be created at each point. In LO BMS, the creation and annihilation operators map to wilson lines, while the eikonal factor corresponds to the decay rates from high energy modes to softer modes. The difficult part though is not finding the form of the equation, but demonstrating factorization, that is, the system is Markovian to leading log. Effective field theory and in particular the soft jet expansion of \Ref{Larkoski:2015zka} give a systematic way to calculate the \emph{resummed} decay rates of the Lindblad equation for jet physics (the BMS equation) and its subleading logarithmic extensions such that the sum over the allowed phase space for the decays is well-behaved in all corners.

 For jet physics, the leading log factorization was accomplished long ago with the ``jet calculus'' of \cite{Marchesini:1983bm,Bassetto:1984ik,Marchesini:1987cf}, however, instead of writing evolution equations for the reduced density matrix\footnote{ But there was a consideration for time-evolution of the wavefunction \cite{Catani:1985xt}.}, the authors sparked the now well-established tradition of Monte-Carlo simultation of jets that properly accounts for soft coherence. These Monte-Carlos solve these evolution equations \cite{Nagy:2007ty}, at least approximately, ultimately the more important endeavor. 

\section{Conclusions}
I have presented a factorization theorem that can be exploited to resum terms contributing to subleading NGLs in jet cross-sections. Moreover the factorization theorem itself will be helpful in understanding the boundary structure of cone algorithms, and have useful applications in jet substrucutre studies. The resummation is distinctly collinear, and captures the fragmentation effects of soft jets at the jet boundary. Using the factorization theorem to calculate soft jet production, I gave a simple extension of the BMS equation that would include the leading collinear effects, but is still not single log accurate. Interestingly, it captures logarithmic contributions that are naively missed by the NLO BMS kernel, but would be corrected when calculating the IR trace of the reduced density matrix as generated by BMS evolution at NLO order. Taken together, the NLO kernel and the collinear resummation, they form a complete NLL exponentiation of the subleading NGLs, with a trivial structure in the IR matrix element.  The soft jet expansion then, which can be systematic up to the number of jets included, provides an important check on what effects are captured in these evolution equations, and can be used to resum the evolution kernels thesmelves. 

With regards to phenomenology, it is plausible that the bulk of these collinear effects at large-$N_c$ are already captured by antennae-dipole showers, like \vincia{} \cite{Giele:2007di,Giele:2011cb} or \ariadne{} \cite{Lonnblad:1992tz}. Given these showers capture leading log soft coherence and parton splitting at each splitting/emission, for all phase space points, they would naturally incorporate many of the subleading logarithms, especially the double collinear logs. However, some terms present in QCD, like the subleading NGLs with rational coefficients, would only be included once the parton shower was matched to a fixed order calculation with sufficient number of loops, or if the shower was constructed around a $2\rightarrow 4$ splitting scheme. Indeed, this suggests that the appropriate construction of NLO BMS equation should be around antennae-multipoles instead of soft currents \cite{Kosower:1997zr,Kosower:2003bh}, to control the decay rates into IR physics. These naturally include both soft coherence and collinear splittings. Then one would be constructing the shower equation \cite{Nagy:2007ty,KOSSAKOWSKI1972247,Lindblad:1975ef,Gorini:1976cm} for a full scale NLL parton shower. 

Perhaps more important is the implications for the B-JIMWLK hierarchy \cite{Mueller:1993rr,Mueller:1994jq,Balitsky:1995ub,Balitsky:1998kc,Balitsky:1998ya,JalilianMarian:1996xn,JalilianMarian:1997gr,Kovchegov:1999yj,Kovner:1999bj,Iancu:2000hn,Iancu:2001ad,Blaizot:2002np}\footnote{For a detailed discussion of the assumptions behind B-JIMWLK, see \cite{Caron-Huot:2013fea}, and for the NLO corrections, see \cite{Balitsky:2008zza,Balitsky:2009xg,Kovner:2013ona}.}. NGLs and small-x physics are known to be related via a conformal mapping \cite{Hatta:2008st,Avsar:2009yb}, including the NLO corrections to B-JIMWLK, \cite{Caron-Huot:2015bja}, whose conformal properties are more delicate \cite{Kovner:2014xia}. This relationship is so-far exact in $N=4$ SYM, and departures from it are directly related to the beta function of QCD. However, both the BFKL \cite{Kuraev:1977fs,Balitsky:1978ic} and the B-JIMWLK hierarchy resummations are noticeably improved at NLO with the inclusion of at least a partial collinear resummation of the evolution kernels \cite{Salam:1998tj,Iancu:2015vea,Iancu:2015joa}. It would be pleasing to exhibit a conformal mapping between the collinear resummation applied here, and the small-$x$ regime, at least for $N=4$ SYM. For the NGL case, the jet radius provided a natural collinear cut-off for these subleading logarithms. Under the stereographic projection that relates the jet physics to small-x, the jet boundary is mapped to a circle in the impact parameter plane. Thus the concern would be collinear splittings between large and small transverse momentum scales (small and large values of the impact parameter). However, in the B-JIMWLK case there is no emergent IR cutoff like that from the jet boundary. In a conformal theory, all subjets produces by B-JIMWLK are then deep in the ``fat jet'', and the collinear improvement seems pointless since there is no IR scale to run to. In QCD, one must worry about $\Lambda_{QCD}$ giving the IR cutoff to the impact parameter space. Given the importance of these collinear improvements, this suggests complete quantification of all participating momentum regions in the high-energy forward scattering regime is lacking. 

Finally, given recent work on resumming all leading logs at finite-$N_c$ in the hemisphere case \cite{Hagiwara:2015bia}, the stage seems set to numerically assess the impact of the NLL NGLs, at least out to moderate values of the non-global logs $\frac{\alpha_s C_A}{\pi}$ln$\frac{m_L}{m_R}\sim 2$. It would be relatively straightforward to work to NLL with two soft subjet contributions, including collinear resummations, and comparing against the collinearly improved LO+NLO BMS, which seems to be adaptable to the methods of \cite{Hagiwara:2015bia}. Like in the small-x case where the collinear improvement allowed the NLO BK evolution to maintain positivy, \Ref{Iancu:2015vea,Iancu:2015joa}, it would be interesting to see if the same obtains in the jet case. 

\begin{acknowledgments}
I wish to especially thank Ian Moult and Andrew Larkoski for colloboration on these and related topics, and many fruitful discussions about NGLs. I also wish to thank Iain Stewart, Richard Holman, and Simon Caron-Huot for helpful discussions. This work is supported by the U.S. Department of Energy (DOE) under grant Contract Numbers DE-SC00012567 and DE-SC0011090, as well as an MIT Pappalardo Fellowship.
\end{acknowledgments}

\appendix
\section{Plus Distributions}\label{Plus_Distros}
I briefly summarize the properties of the plus distributions found in the renormalized momentum space functions. I follow the definition given in \cite{Ligeti:2008ac}. Let $q(x)$ be a function less singular than $1/x^2$ at the origin, then the rgularized plus distribution with boundary point $x_0$ is given by:
\begin{align}
\Big[q(x)\Big]_+^{[x_0]}&=\lim_{\delta\rightarrow 0}\frac{d}{dx}\Big[\theta(x-\delta)Q(x,x_0)\Big]\\
Q(x,x_0)&=\int_{x_0}^xdx' q(x')
\end{align}
Integrating against a function $f(x)$, the above plus distribution yields:
\begin{align}\label{eq:int_plus}
\int_0^{x_{max}}dx'f(x')\Big[q(x')\Big]_+^{[x_0]}=\int_0^{x_{max}}dx' q(x')[f(x')-f(0)]+f(0)Q(x_{max},x_0)
\end{align}
If $x_0=1$, then I adopt the convention:
\begin{align}
\Big[q(x)\Big]_+=\Big[q(x)\Big]_+^{[1]}
\end{align}
Now I can define the distributions:
\begin{align}
{\mathcal L}^a(x)&=\Big[\frac{1}{x^{1-a}}\Big]_+\\
{\mathcal L}_n(x)&=\frac{d^n}{da^n}{\mathcal L}^a(x)\Big|_{a=0}=\Big[\frac{1}{x}\text{ln}^n x\Big]_+\\
{\mathcal L}_a(\lambda,x)&=\frac{1}{\lambda}{\mathcal L}_a\Big(\frac{x}{\lambda}\Big)\\
{\mathcal L}_n(\lambda,x)&=\frac{1}{\lambda}{\mathcal L}_n\Big(\frac{x}{\lambda}\Big)
\end{align}
These last distributions satisfy the rescaling identity:
\begin{align}
{\mathcal L}^a(\lambda,x)&=\lambda^{-a}{\mathcal L}^a(x)+\delta(x)\frac{\lambda^{-a}-1}{a}\\
{\mathcal L}_n(\lambda,x)&=\sum_{k=0}^n \,_nC_k \text{ln}^k\Big(\lambda^{-1}\Big) {\mathcal L}_{n-k}(x)+\delta(x)\frac{\text{ln}^{n+1}\Big(\lambda^{-1}\Big)}{n+1}
\end{align}
This allows us to take the logarithmic derivative:
\begin{align}
\frac{d}{d\text{ln}\lambda}{\mathcal L}_n(\lambda,x)&=\sum_{k=1}^n -k\,_nC_k \text{ln}^{k-1}\Big(\lambda^{-1}\Big) {\mathcal L}_{n-k}(x)-\delta(x)\text{ln}^{n}\Big(\lambda^{-1}\Big)\\
\frac{d}{d\text{ln}\lambda}{\mathcal L}_0(\lambda,x)&=-\delta(x)\\
\frac{d}{d\text{ln}\lambda}{\mathcal L}_1(\lambda,x)&=\delta(x)\text{ln}\lambda-{\mathcal L}_0(\lambda,x)
\end{align}
Finally, the plus distributions encountered in this paper have the following laplace transforms:
\begin{align}
\int_{0}^{\infty} dx {\mathcal L}^a(\lambda,x) e^{-x\,\tau}&=\int_{0}^{\infty} dx\frac{\lambda^{-a}}{x^{1-a}}\Big(e^{-x\,\tilde\tau}-1\Big)+\int_{1}^{\infty} dx\frac{\lambda^{-a}}{x^{1-a}}+\frac{\lambda^{-a}-1}{a}\nonumber\\
&=\big(\lambda\tau\big)^a\,\Gamma(a)-\frac{1}{a}\\
\int_{0}^{\infty} dx {\mathcal L}_0(\lambda,x) e^{-x\,\tau}&=-\text{ln}\Big(\tau\lambda e^{\gamma_E}\Big)\\
\int_{0}^{\infty} dx {\mathcal L}_1(\lambda,x) e^{-x\,\tau}&=\frac{1}{2}\text{ln}^2\Big(\tau\lambda e^{\gamma_E}\Big)+\frac{\pi^2}{12}
\end{align}

\section{Energy Correlation Functions}\label{app:ecfs}
The energy correlation functions are defined as:
\begin{align}\label{eq:ecf_def}
\ecf{2}{\alpha}&= \frac{1}{E_J^2} \sum_{i<j\in J} E_i E_j \left(
\frac{2p_i \cdot p_j}{E_i E_j}
\right)^{\alpha/2} \,, \\
\ecf{3}{\alpha}&= \frac{1}{E_J^3} \sum_{i<j<k\in J} E_i E_j E_k \left(
\frac{2p_i \cdot p_j}{E_i E_j}
\frac{2p_i \cdot p_k}{E_i E_k}
\frac{2p_j \cdot p_k}{E_j E_k}
\right)^{\alpha/2} \,. \nonumber
\end{align}
Here $J$ denotes the jet,  $p_i$ and are the four momentum and energy of particle $i$ in the jet. 

\section{Out of Jet Anomalous Dimensions For BMS}\label{Out_Of_Jet_Anom}
I give the explicit forms in the hemisphere case of the out-of-jet anomalous dimension of \Eq{eq:BMS_anom_dim_ab}, as calculated in \cite{Schwartz:2014wha}. If $a,b$ are light-like vectors within the fat jet, we have:
\begin{align}
S_{ab}(j)&=-\text{ln}\Big(\text{cos }\theta_j\Big)-\frac{1}{2}\text{ln}\Big(\frac{[ab]}{2[aj][jb]}\Big)\\
&=-\text{ln}\Big(1-\text{tan}^2\frac{\theta_j}{2}\Big)-\text{ln}\Big(\frac{1+\text{cos }\theta_j}{2}\Big)-\frac{1}{2}\text{ln}\Big(\frac{[ab]}{2[aj][jb]}\Big)
\end{align}
When $b=\nbar$, we have:
\begin{align}
S_{a\nbar}(j)&=-\text{ln}\Big(\text{cos }\theta_j\Big)-\frac{1}{2}\text{ln}\Big(\frac{(a\nbar)}{2[aj](j\nbar)}\Big)\\
&=-\text{ln}\Big(1-\text{tan}^2\frac{\theta_j}{2}\Big)-\text{ln}\Big(\frac{1+\text{cos }\theta_j}{2}\Big)-\frac{1}{2}\text{ln}\Big(\frac{[ab]}{2[aj][jb]}\Big)
\end{align}
Note: $\text{tan}^2\frac{\theta_j}{2}=\frac{n\cdot n_j}{\bar\cdot n_j}\equiv\frac{u}{v}$. I define the angular products:
\begin{align}
(ab)&=1-\text{cos}\theta_{ab}=1-\text{cos}\theta_a\text{cos}\theta_b-\text{cos}(\phi_a-\phi_b)\text{sin}\theta_a\text{sin}\theta_a\\
[ab]&=(\bar {a} b)=1+\text{cos}\theta_a\text{cos}\theta_b-\text{cos}(\phi_a-\phi_b)\text{sin}\theta_a\text{sin}\theta_a
\end{align}
$\theta_{ab}$ is the angle between the null vectors, and $\bar{a}$ is the reflection of the spatial components of $a$ into the other hemisphere. We have given the explicit forms in the spherical coordinates defined by the fat jet axis $\hat{n}$.

\section{Joint Resummation of Soft Jets }\label{app:threshold_soft_jet}
To incorporate the collinear double logs into the BMS equation (\Sec{sec:collinear_BMS}), we need to combine the factorization of this paper, with that of \Ref{Larkoski:2015zka}. Thus we want a power counting scheme:
\begin{align}\label{eq:Joint_resum_phase_space}
u\gg \outj \gg \outj\Big(1-\frac{u}{v}\text{cot}^2\frac{R}{2}\Big)\,.
\end{align}
First we recall the factorization of \Ref{Larkoski:2015zka}:
\begin{align}
\label{eq:SJ_fact}
\frac{d\sigma}{d\ecf{2}{\alpha}\,d\ecf{2}{\beta}\,d\ecf{3}{\alpha}d\outj}&=\sigma_0 H(Q^2)H_{n\bar{n}}^{sj}\Big(\ecf{2}{\alpha},\ecf{2}{\beta},\outj\Big)S_{n\nbar \sja}\Big(\ecf{3}{\alpha};\outj;R\Big)\otimes S_{\sja\sjabar}(\ecf{3}{\alpha},R)\nonumber\\
&\qquad\otimes\, J_{\sja}(\ecf{3}{\alpha},Q_{sj})\otimes J_{n}(\ecf{3}{\alpha})\otimes J_{\nbar}(\outj)
\end{align}
In the region of \Eq{eq:Joint_resum_phase_space} of phase space, the soft jet factorization theorem takes the form:
\begin{align}
\label{eq:Threshold_fact}
\frac{d\sigma}{d\ecf{2}{\alpha}\,d\ecf{2}{\beta}\,d\ecf{3}{\alpha}d\outj}&=\sigma_0 H(Q^2)H_{n\bar{n}}^{sj}\Big(\ecf{2}{\alpha},\ecf{2}{\beta},\outj\Big)H_{n\nbar \sja}\Big(\ecf{3}{\alpha};\outj;R\Big)C_{\sja}^{\text{thr}}(\outj,R)\nonumber\\
&\qquad\otimes S_{\sja\sjabar}(\ecf{3}{\alpha},R)\otimes\, {\mathcal E}_{\sja}(\ecf{3}{\alpha},Q_{sj})\otimes J_{n}(\ecf{3}{\alpha})\otimes J_{\nbar}(\outj)\otimes S_{n\nbar}\Big(\ecf{3}{\alpha};\outj;R\Big)
\end{align}
The matching coefficient $C_{\sja}^{\text{thr}}(\outj,R)$ is related to that of \Sec{sec:matching_edge_of_jet_function} with the added expansion $v\gg \outj$. This up to constants drops all terms but the double logs in \Eq{eq:Edge_Of_Jet_fin_in_out}. We have converted the soft function into a hard function $S_{n\nbar \sja}$ by subtracting the contribution of $C_{\sja}^{\text{thr}}(\outj,R)$:
\begin{align}\label{eq:Soft_Function_matching_for_threshold}
S_{n\nbar \sja}\Big(\ecf{3}{\alpha};\outj;R\Big)&=H_{n\nbar \sja}\Big(\ecf{3}{\alpha};\outj;R\Big)C_{\sja}^{\text{thr}}(\outj,R)\otimes S_{n\nbar}\Big(\ecf{3}{\alpha};\outj;R\Big)+...
\end{align}
This localizes the scale $1-\frac{u}{v}\text{cot}^2\frac{R}{2}$ into the function $C_{\sja}^{\text{thr}}$, and the scale $\outj$ into the hard matching coefficent. The new three wilson line soft function is equivalent that found in \Eq{eq:tripole_softs}. To get the resummation of the double collinear logs, we evolve $C_{\sja}^{\text{thr}}$ between the scales:
\begin{align}
\outj^c\Big(1-\frac{u}{v}\text{cot}^2\frac{R}{2}\Big) \rightarrow \outj^c\,.
\end{align}
To leading log in the double collinear logs, we can simply exponentiate the logs, but at next-to-leading log and beyond, a more complicated resummation factor like \Eq{eq:dl_cumulant_evo} may be necessary in general. However, in a conformal theory, performing the resummation in laplace space with canonical scale setting, the dependence on the out-of-jet scale vanishes completely in the resummation, so that the exponentiated form of the double logs would hold to all orders, with only corrections to the anomalous dimension. To finish the resummation, we run $H_{n\nbar \sja}$ and $C_{\sja}^{\text{thr}}$ together to the scale $\mu$ from $\outj^c$. From the matching equation \eqref{eq:Soft_Function_matching_for_threshold}, and appropriately reassociating the functions of the factorization theorem to cancel global effects, this is just the running of the $G_{n\nbar\sja}$ factor of \Ref{Larkoski:2015zka}. Together, these resummations give the form \Eq{eq:collinear_improved_U_factor}.

Explicitly, the evolution proceeds as follows. First I take the matching for arbitrary dipole $a, b$ radiating a soft jet:
\begin{align}\label{eq:Soft_Function_matching_for_threshold_arb}
S_{ab \sja}\Big(\ecf{3}{\alpha};\outj;R\Big)&=H_{ab \sja}\Big(\ecf{3}{\alpha};\outj;R\Big)C_{\sja}^{\text{thr}}(\outj,R)\otimes S_{ab}\Big(\ecf{3}{\alpha};\outj;R\Big)+...
\end{align}
Take the edge of jet anomalous dimension at threshold:
 \begin{align}
\mu\frac{d}{d\mu}\text{ln}C_{\sja}^{\text{thr}}(\outjlp,R)S_{n\nbar \sja}&=-\frac{\alpha_s(\mu) C_A}{\pi}\Big(\text{ln}(\mu\outjlp e^{\gamma_E})-\text{ln}\Big(1-\frac{n\cdot\,\sja}{n\cdot\,\sjabar}\text{cot}^2\frac{R}{2}\Big)\Big)
\end{align}
I evolve between the scales:
\begin{align}
\outj^c\Big(1-\frac{n\cdot\,\sja}{n\cdot\,\sjabar}\text{cot}^2\frac{R}{2}\Big)\rightarrow \outj^c\Big(1-\frac{n\cdot\,\sja}{n\cdot\,\sjabar}\text{cot}^2\frac{R}{2}\Big)e^{S_{ab}(\sja)}
\end{align}
Though awkward, from the explicit expressions for the out of jet anomalous dimensions, $S_{ab}(\sja)$ in \Sec{Out_Of_Jet_Anom}, this last scale does not actually depend on the angle to the jet boundary. This intermediate scale is chosen as the scale where the out-of-jet component of the anomalous dimension of $H_{a b\sja}$ vanishes in \eqref{eq:Soft_Function_matching_for_threshold_arb}, so the matching logs are minimized. I then evolve using the standard dressed gluon anomalous dimension \eqref{eq:BMS_anom_dim_ab} between the scales:
\begin{align}
\outj^c\Big(1-\frac{n\cdot\,\sja}{n\cdot\,\sjabar}\text{cot}^2\frac{R}{2}\Big)e^{S_{ab}(\sja)}\rightarrow \mu
\end{align}
I, up to running coupling effects that are easily included, obtain \Eq{eq:collinear_improved_U_factor_arb_dipole}. The appearance of $\outj^c\Big(1-\frac{n\cdot\,\sja}{n\cdot\,\sjabar}\text{cot}^2\frac{R}{2}\Big)e^{S_{ab}(\sja)}$ in the intermediate scale is just a consequence of the function $H_{a b\sja}$ in \eqref{eq:Soft_Function_matching_for_threshold_arb} being sensitive to positions of the directions $a,b$ within the fat jet.

\section{In-In Contribution To Edge-of-Jet Function}
The in-in contribution can be written in terms of a standard jet function contribution with no jet boundary contributions, and the jet boundary contributions:
\begin{align}
{\mathcal E}_{\sja}^{(1)}\Big(\ecflp{3}{\alpha};\outj;Q_{sj};R\Big)\Bigg|_{\text{in-in}}&=\delta(\outj)\Big({\mathcal E}_{\sja}^{(1)}\Big(\ecflp{3}{\alpha};Q_{sj}\Big)\Big|_{\text{stnd.}}+\delta {\mathcal E}_{\sja}^{(1)}\Big(\ecflp{3}{\alpha};Q_{sj};R\Big)\Big)
\end{align}
The standard contribution is given by:
\begin{align}\label{eq:edge_of_jet_in_in_stnd_start}
{\mathcal E}_{\sja}^{(1)div}\Big(\ecflp{3}{\alpha};Q_{sj}\Big)&=\frac{\alpha_sC_A}{\pi}\Bigg\{\frac{\alpha}{2(1-\alpha)\epsilon^2}+\frac{\beta_0}{\epsilon}+\frac{\alpha}{(1-\alpha)\epsilon}\text{ln}\Bigg(F\Bigg(\frac{\mu}{Q_{sj}}\ecflp{3}{\alpha},\frac{Q_{sj}}{Q};N\Bigg)\Bigg)\Bigg\}\\
{\mathcal E}_{\sja}^{(1)fin}\Big(\ecflp{3}{\alpha};Q_{sj}\Big)&=\frac{\alpha_sC_A}{\pi}\Bigg\{C^{(1)}+\text{ln}\Bigg(F\Bigg(\frac{\mu}{Q_{sj}}\ecflp{3}{\alpha},\frac{Q_{sj}}{Q};N\Bigg)\Bigg)\nonumber\\
&\qquad\qquad\Bigg(\frac{\beta_0}{C_A}+\frac{\alpha}{1-\alpha}\text{ln}\Bigg(F\Bigg(\frac{\mu}{Q_{sj}}\ecflp{3}{\alpha},\frac{Q_{sj}}{Q};N\Bigg)\Bigg)\Bigg)\Bigg\}\\
C^{(1)}&=\frac{67(1-\alpha)}{18\alpha}-\frac{\pi^2}{24}\frac{9\alpha^2-16\alpha+4}{(1-\alpha)\alpha}+n_fT_f\frac{23\alpha-26}{36 C_A \alpha}\\
F\Bigg(\frac{\mu}{Q_{sj}}\ecflp{3}{\alpha},\frac{Q_{sj}}{Q};N\Bigg)&=e^{\gamma_E/\alpha}N^{\frac{1}{\alpha}}\Big(\frac{Q_{sj}}{Q}\Big)^{\frac{2}{\alpha}}\frac{\mu}{Q_{sj}}\Big(\ecflp{3}{\alpha}\Big)^{\frac{1}{\alpha}}
\end{align}
The boundary dependent pieces are given as:
\begin{align}
\delta {\mathcal E}_{\sja}^{(1)div}\Big(\ecflp{3}{\alpha};Q_{sj};R\Big)&=-\frac{C_A\alpha_s}{\pi}\Bigg(\frac{1}{2\epsilon^2(1-\alpha)}+\frac{1}{\epsilon(1-\alpha)}\text{ln}\Bigg(G\Big(\ecflp{3}{\alpha};Q_{sj};\sja;R\Big)\Bigg)\\
\delta {\mathcal E}_{\sja}^{(1)fin}\Big(\ecflp{3}{\alpha};Q_{sj};R\Big)&=\frac{C_A\alpha_s}{\pi}\Bigg\{-\frac{1}{2(1-\alpha)}\text{ln}^2\Bigg(G\Big(\ecflp{3}{\alpha};Q_{sj};\sja;R\Big)\Bigg)\nonumber\\
&\qquad\qquad-\frac{\pi^2(2\alpha^2-6\alpha+7)}{12(1-\alpha)}\Bigg\}\\
G\Big(\ecflp{3}{\alpha};Q_{sj};\sja;R\Big)&=2^{1-\alpha}e^{\gamma_E}N\frac{\mu}{Q}\frac{ Q_{sj}}{Q}\ecflp{3}{\alpha}\Big(\text{tan}^2\frac{R}{2}-\frac{n\cdot\sja}{\nbar\cdot\sja}\Big)^{\alpha-1}\Big(\frac{n\cdot\sja}{\nbar\cdot \sja}\Big)^{\frac{1-\alpha}{2}}(\nbar\cdot\sja)^{\alpha-1}\label{eq:edge_of_jet_in_in_bsoft_fin}
\end{align}
These boundary dependent pieces have finite corrections that vanish as $\ecf{3}{\alpha}\rightarrow 0$. The leading terms given here of the boundary dependent contribution can be factored into a boundary soft function.

\bibliography{ngl_factorization}

\end{document}